\documentclass[11pt,a4paper]{article}

\usepackage[margin=1in]{geometry}

\usepackage{amsthm}
\usepackage{amsmath}

\usepackage{newtxtext}
\usepackage{newtxmath}

\usepackage{booktabs}
\usepackage{tabularx}
\usepackage{longtable}
\usepackage{array}
\newcolumntype{Y}{>{\raggedright\arraybackslash}X}

\usepackage{setspace}
\usepackage{authblk}
\usepackage{natbib}
\setcitestyle{authoryear,round}

\usepackage[english]{babel}
\usepackage{graphicx}
\usepackage{float}
\usepackage{placeins}
\usepackage{needspace}
\usepackage{subcaption}

\usepackage[shortlabels]{enumitem}
\usepackage{hyperref}

\makeatletter

\newcommand{\Rmnum}[1]{\expandafter\@slowromancap\romannumeral #1@}
\makeatother

\theoremstyle{definition}

\title{Rachev-Ratio Portfolio Optimization with TimeGAN Scenarios: Replacement versus Supplementation}

\author{Yifan He\thanks{Email: \href{mailto:Yifan.He@ttu.edu}{Yifan.He@ttu.edu}}}

\affil{\small Department of Mathematics and Statistics, Texas Tech University, Lubbock, TX 79409, USA}

\begin{document}

\date{}
\maketitle

\begin{abstract}
\noindent This study evaluates the reliability and investment consequences of scenarios from a specified time-series generative adversarial network (TimeGAN) in Rachev-ratio portfolio allocation. Using 12 exchange-traded funds, we compare historical, replacement and hybrid policies across five training seeds with delayed execution and transaction costs. Synthetic probability mass varies independently of draw count under a fixed training protocol; optimal generator fidelity is not assumed. Generated scenarios understate volatility and tail losses and distort dependence. Limited longer-training diagnostics improve reconstruction loss without improving cross-asset correlation accuracy. Higher-mass hybrids earn larger average net returns in some configurations, but all conditional generated-versus-historical return intervals include zero at the reference block length. Full-period TimeGAN means exceed resampling controls, while direct paired return, Sharpe and realized-tail-ratio intervals include zero. Replacement raises turnover and is sensitive to the modified objective's denominator floor; changing that floor does not uniformly improve realized downside risk. Exploratory generation-error associations do not identify a causal investment-performance channel. The results illustrate financial model risk from the interaction of imperfect scenarios and a tail-sensitive allocation criterion, supporting joint evaluation of fidelity, numerical behavior and realized outcomes within the tested implementation.\par\medskip\noindent
\end{abstract}

\section{Introduction}\label{sec:introduction}

Portfolio optimization depends on both the objective and the scenario distribution supplied to it. Historical windows contain relatively few extreme observations, motivating the use of learned generators to simulate additional return vectors. A generator may nevertheless reproduce central features while distorting the tails or dependence relevant to a portfolio decision. Statistical resemblance, the number of simulated rows, the attained in-window objective and realized investment performance are consequently distinct objects of evaluation.

This study asks whether time-series generative adversarial network (TimeGAN) scenarios provide useful inputs to an implemented Rachev ratio (RR) allocation rule. The procedures are the historical Rachev-ratio portfolio (RRP), the artificial-intelligence-based Rachev-ratio portfolio (AI-RRP) using replacement, and the hybrid Rachev-ratio portfolio (Hybrid-RRP) using supplementation. Replacement discards the historical empirical distribution after fitting; supplementation preserves historical observations with a specified probability mass. We separate synthetic probability mass from draw count, reuse fitted networks across configurations, and assess both scenario fidelity and subsequent net returns. The comparison is conditional on the selected generator, numerical procedure and exchange-traded fund (ETF) universe.

The objective is not to identify the best-performing TimeGAN architecture or the maximum attainable fidelity of synthetic financial data. Instead, we ask how scenarios produced by a specified generative implementation affect a tail-sensitive allocation rule when their distributional fidelity is imperfect. Fidelity and downstream decision quality are separate properties: incomplete statistical resemblance does not by itself determine investment usefulness, and a favorable optimization score need not imply favorable realized performance. The observed distortions therefore motivate evaluating financial model risk through the joint evidence on scenarios, numerical behavior and investment outcomes. The experiment was not designed as a deliberate low-fidelity stress test; this interpretation follows the recorded diagnostics and does not establish a causal error-to-performance mechanism.

The completed experiment fits 198 historical windows under each of five training seeds and generates three independent draws per fit. The allocation design comprises four hybrid mass--count cells and two replacement counts for each seed, plus a common historical reference. All 31 policies use the same fractional-tail convention, execution dates, costs and numerical candidate-selection rule. Equal-weight rebalancing and buy-and-hold provide additional deterministic references. A separate 62-policy full-period calculation compares denominator floors and simple historical resampling using the same 198 decision dates and original TimeGAN draws. Baselines and sensitivity arms are recomputed in one runtime, with independently verified allocations, costs and paired conditional inference; these replays do not constitute independent market replications. An exploratory panel analysis also links the saved generation diagnostics to allocations and subsequent outcomes at all 198 decision dates, with explicit limits on causal and statistical interpretation.

The principal findings are mixed. Under the recorded training budget, generated dependence and tails remain inaccurate. Equal-mass hybrids have lower mean net returns than the historical reference; higher synthetic mass produces larger means in some cells, with substantial seed variation. Conditional inference does not establish a stable advantage at the reference block length. Larger synthetic samples do not yield uniform gains across configurations. Numerical cross-checks also reveal that the absolute-value denominator with a small floor admits very large in-window ratios near a zero lower-tail mean. These are features of the implemented objective, not evidence of exceptional investment opportunities.

The contribution is a controlled separation of scenario use, synthetic mass and draw count, together with an auditable evaluation of their interaction with the implemented tail-ratio criterion. Distributional diagnostics alone do not establish that interaction: denominator sensitivity, allocation behavior, costs and realized outcomes supply the downstream evidence. The evaluation is exploratory. 

Appendix~\ref{sec:appendix_results} provides auxiliary distributional, training, allocation and interval-sensitivity diagnostics omitted from the main figures.

The rest of the paper is organized as follows. Section~\ref{sec:literature} reviews related literature. Section~\ref{sec:data} describes the data and execution. Section~\ref{sec:methods} defines the portfolio criteria; Sections~\ref{sec:generation} and~\ref{sec:numerics} examine generation and numerical reliability before the investment results in Section~\ref{sec:investment_results}. Section~\ref{sec:robustness} examines denominator sensitivity, resampling controls and the scope of inference, and Section~\ref{sec:discussion_conclusion} concludes.

\section{Literature Review}\label{sec:literature}
Mean--variance allocation~\citep{markowitz1952portfolio}, conditional value-at-risk (CVaR) optimization~\citep{rockafellar2000optimization} and reward-risk criteria~\citep{biglova2004different,stoyanov2007optimal,farinelli2008beyond,rachev2008advanced} place different demands on estimated distributions. A favorable-to-unfavorable tail ratio is especially sensitive to how both extreme regions are represented. Its improvement need not imply lower absolute losses or higher geometric growth. Passive diversification remains an essential reference given the estimation burden of optimized rules~\citep{demiguel2009optimal}.

Generative adversarial networks (GANs)~\citep{goodfellow2014generative} and TimeGAN~\citep{yoon2019time} motivate learning financial scenario distributions; Quant GANs emphasize features such as volatility clustering~\citep{wiese2020quant}. Downstream uses differ substantially. \citet{koshiyama2021generative} study trading-strategy fine-tuning and combination. \citet{pena2024modified} use modified conditional tabular generative adversarial network (CTGAN) inputs to maximize expected return subject to a CVaR limit, rather than maximize a tail ratio; they use spread-based cost estimates, a five-year lookback and annual decisions. \citet{lu2025structured} evaluate structured-noise generation with a proximal policy optimization trading agent, a fixed historical test period and trading costs of 5 basis points (bp) of traded notional. Recent preprints examine fidelity and portfolio applications~\citep{hounwanou2025financial,hounwanou2025applications}. These studies motivate task-specific evaluation; none establishes that a fitted generator automatically improves the criterion considered here. The relevant differences concern the generator, portfolio task, scenario use, costs and rolling evaluation.

\citet{hounwanou2025financial} compare TimeGAN, variational autoencoders, and autoregressive integrated moving-average models with generalized autoregressive conditional heteroskedasticity in downstream portfolio allocation on a fixed test split. \citet{hounwanou2025applications} examine TimeGAN and variational autoencoders for synthetic-input mean--variance allocation and report a five-year/six-month rolling evaluation. Neither cited version specifies a transaction-cost rule for its reported portfolio comparison; this omission is not treated as a zero-cost specification. The preprints have distinct arXiv identifiers and versions. Unlike these designs, the present experiment separates replacement from supplementation and synthetic probability mass from draw count under one modified tail-ratio allocation procedure, with a 756-day rolling fit, 20-day decisions, and explicit trading costs.

\section{Data and Experimental Design}\label{sec:data}

\subsection{Data and Asset Universe}\label{sec:data_source}

The fixed dataset contains daily split- and distribution-adjusted closing prices from April 11, 2007 to December 31, 2025. The 12 ETFs in Table~\ref{tab:etf_data} span equities, bonds, commodities, and real assets. Prices were obtained through yfinance, and the three-month United States (U.S.) Treasury constant-maturity yield (Federal Reserve series DGS3MO) supplies the risk-free proxy. The frozen workbook contains 4,713 common price dates and 4,712 simple returns. Price and yield calendars are checked before calculation; no forward-looking feature enters the rolling window.

\begin{table}[htbp]
\centering
\small
\caption{Summary of the 12 selected ETFs, ordered alphabetically by ticker symbol}
\label{tab:etf_data}

\setlength{\tabcolsep}{4pt}
\renewcommand{\arraystretch}{1.15}

\begin{tabularx}{\textwidth}{@{}l Y Y c@{}}
\toprule
Ticker & Name & Category & Inception year \\
\midrule

DBC & Invesco DB Commodity Index Tracking Fund
    & Commodities and real assets & 2006 \\

EEM & iShares MSCI Emerging Markets ETF
    & International equities & 2003 \\

EFA & iShares MSCI EAFE ETF
    & International equities & 2001 \\

GLD & SPDR Gold Shares
    & Commodities and real assets & 2004 \\

HYG & iShares iBoxx High Yield Corporate Bond ETF
    & Corporate bonds & 2007 \\

IEF & iShares 7--10 Year Treasury Bond ETF
    & U.S. Treasury bonds & 2002 \\

IWM & iShares Russell 2000 ETF
    & U.S. equities & 2000 \\

LQD & iShares iBoxx Investment Grade Corporate Bond ETF
    & Corporate bonds & 2002 \\

QQQ & Invesco QQQ Trust
    & U.S. equities & 1999 \\

SPY & SPDR S\&P 500 ETF Trust
    & U.S. equities & 1993 \\

TLT & iShares 20+ Year Treasury Bond ETF
    & U.S. Treasury bonds & 2002 \\

USO & United States Oil Fund LP
    & Commodities and real assets & 2006 \\
\bottomrule
\end{tabularx}
\end{table}

The ETF universe is selected retrospectively from funds with a common history, which limits generalization; it is not an ex ante security-selection rule. Section~\ref{sec:data_diagnostics} provides distributional and extreme-date diagnostics.

\FloatBarrier

\subsection{Price Trajectories and Distributional Diagnostics}\label{sec:data_diagnostics}
Appendix Figure~\ref{fig:supp_distribution} reports skewness and excess kurtosis for 4,712 asset returns. All 12 assets have positive excess kurtosis, ranging from 2.60 for IEF to 56.37 for LQD, indicating departures from a Gaussian reference even when skewness is not significant. The label not significant (n.s.) denotes a test result above the stated significance threshold. Dependence and conditional heteroskedasticity limit their conventional independent-observation significance interpretation. The rejection of a Gaussian reference does not justify a particular portfolio objective.

\begin{figure}[!htbp] 
\centering
\includegraphics[width=.78\textwidth]{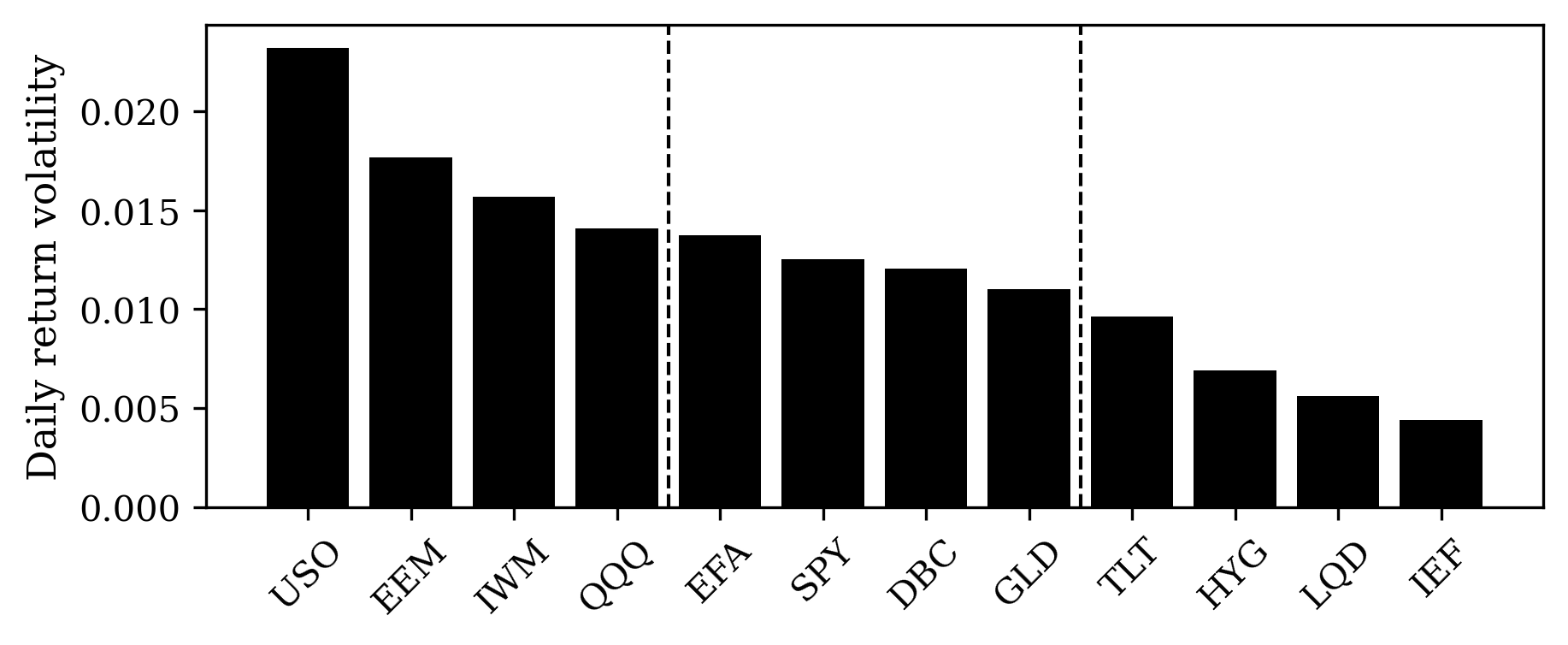}
\caption{Return volatility of the 12 selected ETFs, ranked from highest to lowest}
\label{fig:vol_bar}
\end{figure}

\begin{figure}[p]
\centering
\includegraphics[width=.82\textwidth]{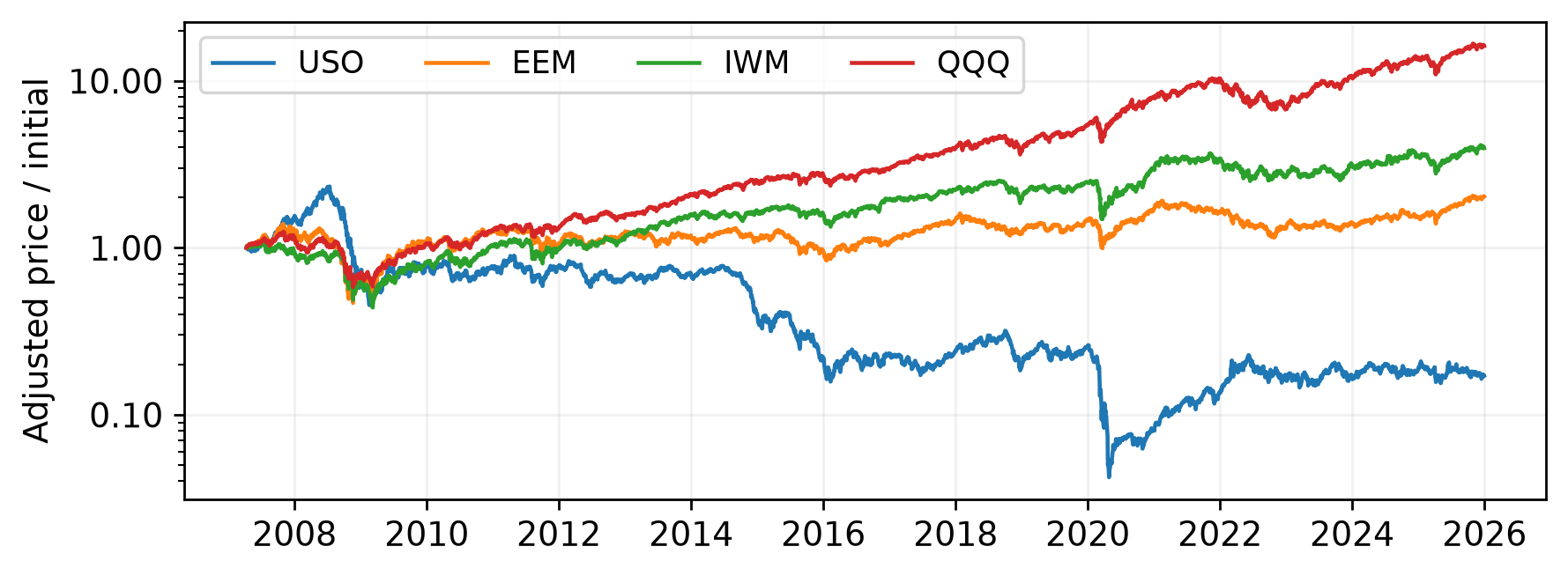}\\[4pt]
\includegraphics[width=.82\textwidth]{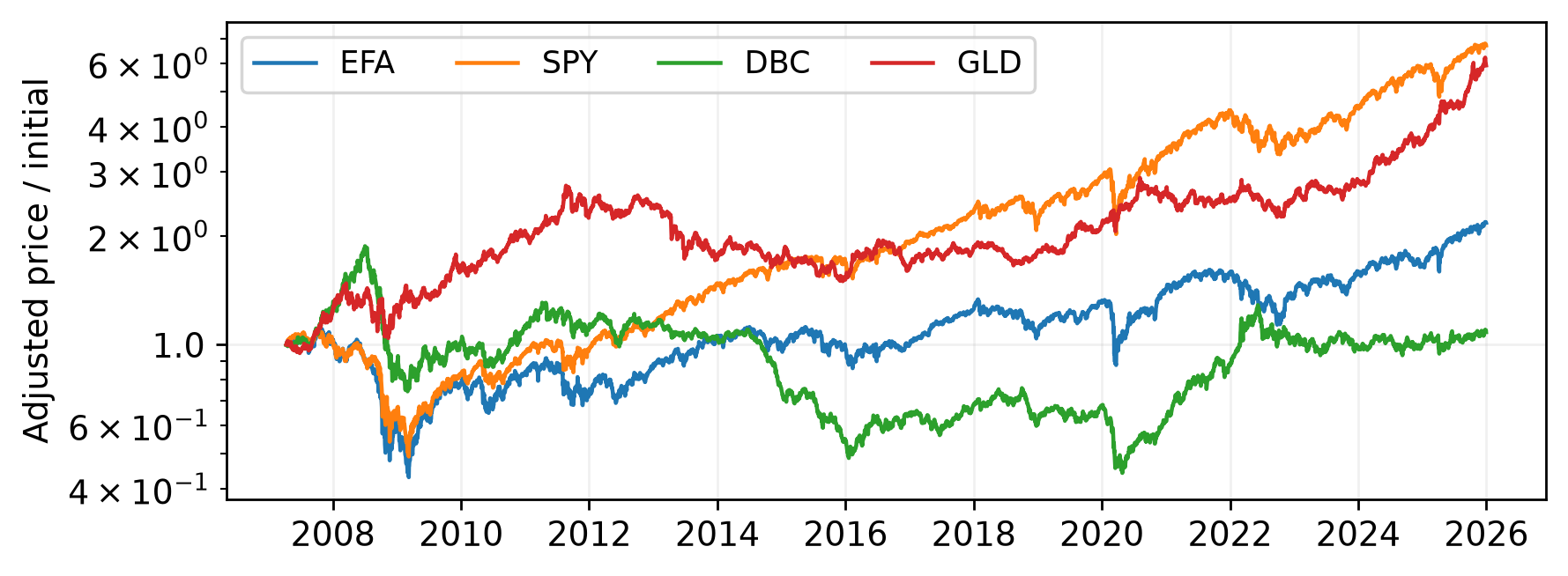}\\[4pt]
\includegraphics[width=.82\textwidth]{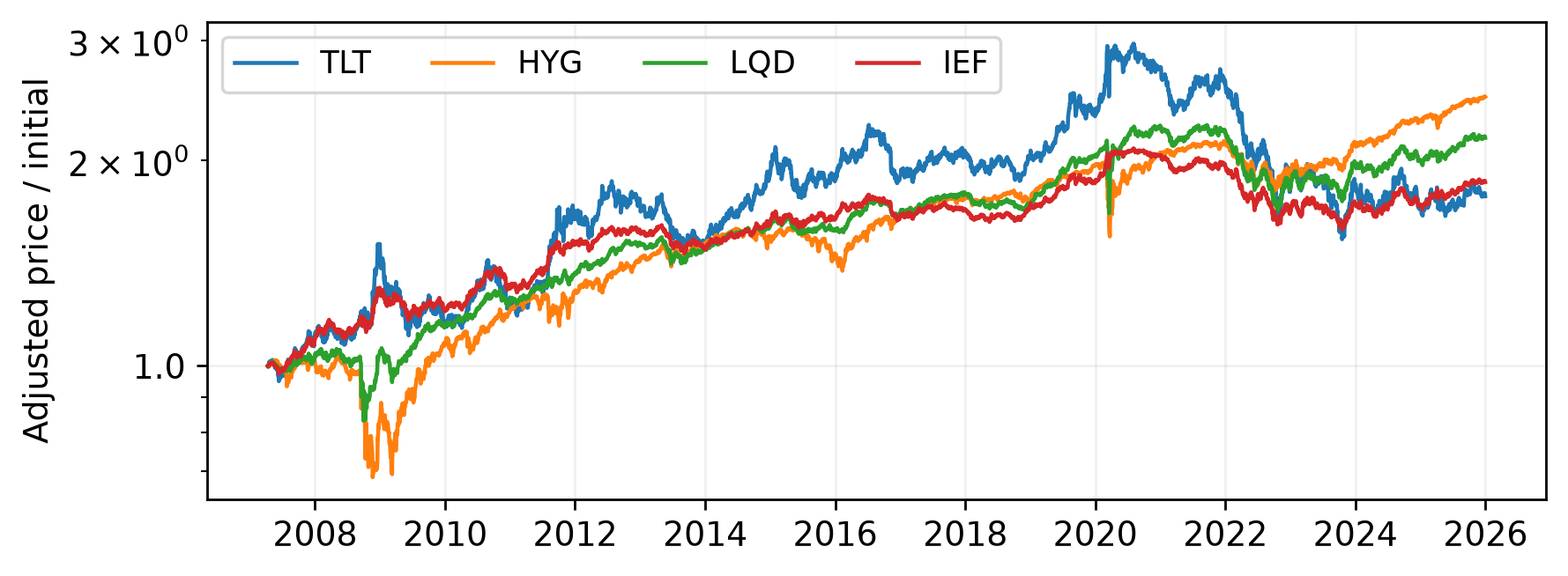}
\caption{Normalized adjusted-price trajectories of the 12 ETFs, grouped by daily return volatility. Each series begins at one on April 11, 2007}
\label{fig:price_etfs}
\end{figure}

\begin{figure}[!htbp]
\centering
\begin{subfigure}{0.80\textwidth}
\includegraphics[width=\linewidth]{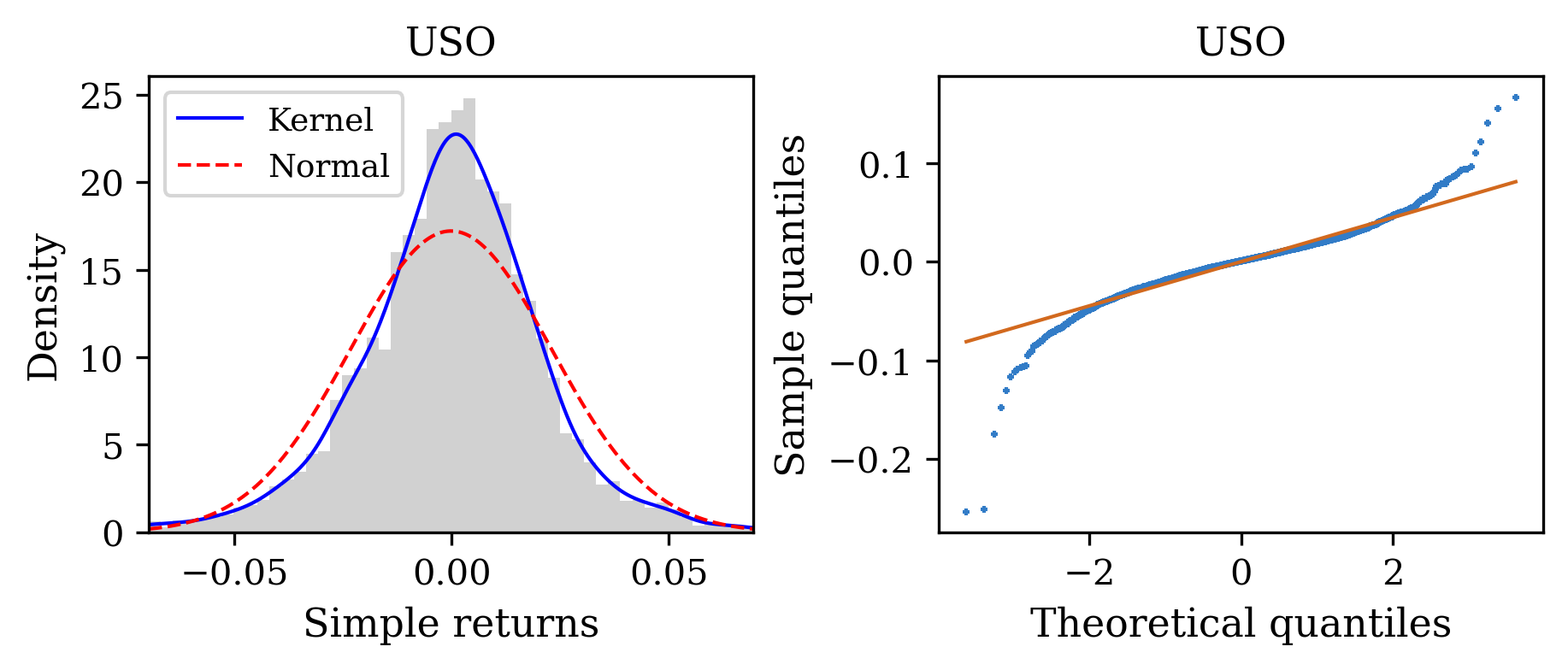}
\caption{USO}
\end{subfigure}\\[8pt]
\begin{subfigure}{0.80\textwidth}
\includegraphics[width=\linewidth]{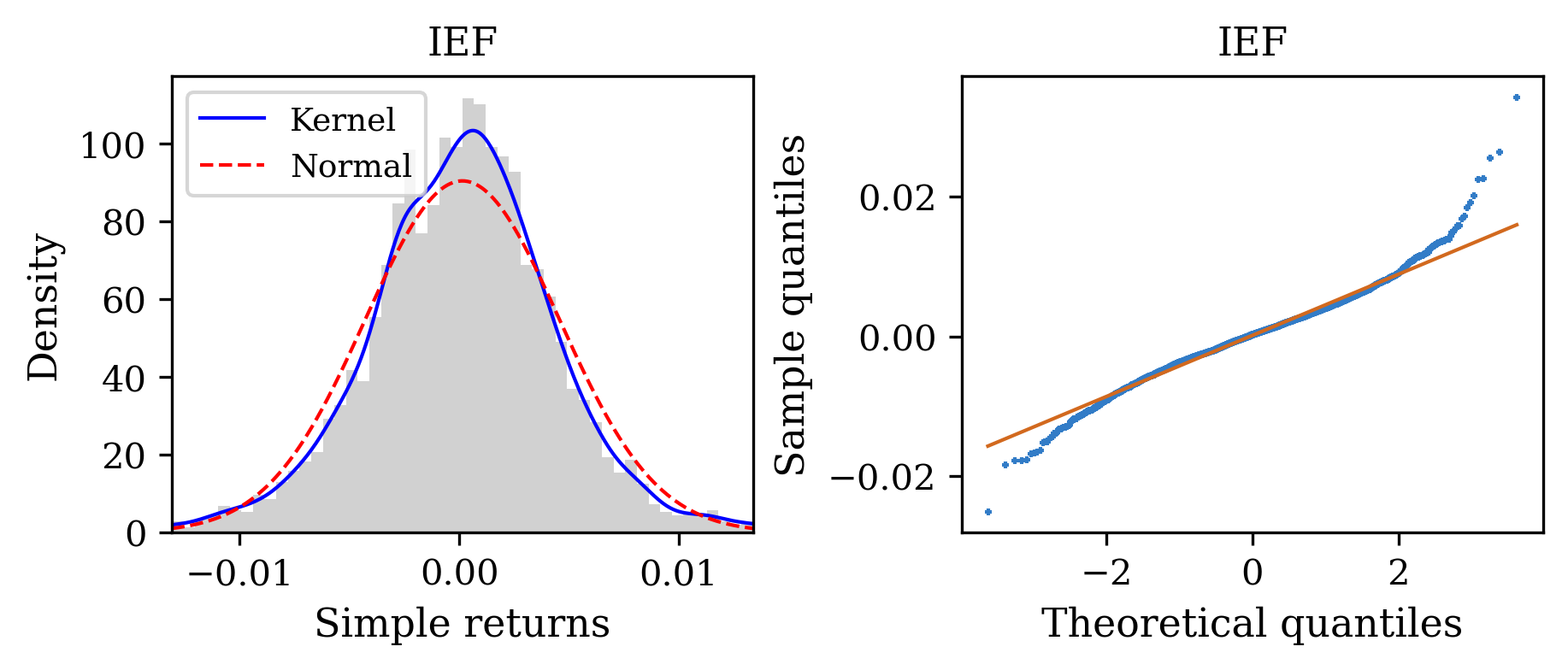}
\caption{IEF}
\end{subfigure}
\caption{Distributional diagnostics for daily USO and IEF returns, the highest- and lowest-volatility assets in the full sample. Each panel pairs a histogram and density comparison with a normal QQ plot}
\label{fig:hist_qq}
\end{figure}

Figure~\ref{fig:vol_bar} ranks the 12 ETFs by their full-sample daily return volatility. The panels in Figure~\ref{fig:price_etfs} retain all 12 normalized price trajectories, grouped by that ranking: QQQ and SPY show substantial long-run appreciation, whereas USO experiences a sustained decline. These contrasting paths make clear that higher daily volatility does not imply greater cumulative price growth. USO has the highest daily volatility (2.32\%) and IEF the lowest (0.44\%). We therefore choose USO and IEF for the normal-reference quantile--quantile (QQ) plots in Figure~\ref{fig:hist_qq}: they illustrate the high- and low-volatility ends of this universe. In Figure~\ref{fig:hist_qq}, both assets deviate from the normal-reference line in their tails; USO also displays a particularly extreme negative tail. Thus, lower volatility does not imply Gaussian tails. This is a descriptive full-sample selection, not an asset-selection rule used by the rolling optimizer; two endpoint examples do not establish the shape of every asset distribution.

Twenty extreme LQD/HYG dates were cross-checked with independently published issuer NAVs and distributions; Appendix Figure~\ref{fig:supp_market_nav} shows six illustrative cases from these checks. For LQD on September 30, 2008, the adjusted market return is 9.77\% while the distribution-adjusted NAV return is 0.13\%, illustrating that extreme market-price movements need not be mirrored by NAV movements. NAV and adjusted market closing price are different quantities; the check neither proves a data error nor independently verifies the exact closing-price observations. No return is removed. Issuer material describes ETF market-price/NAV differences during stressed trading~\citep{blackrock2020pricing}.

\begin{figure}[!htbp] 
	\centering
	\includegraphics[width=.80\textwidth]{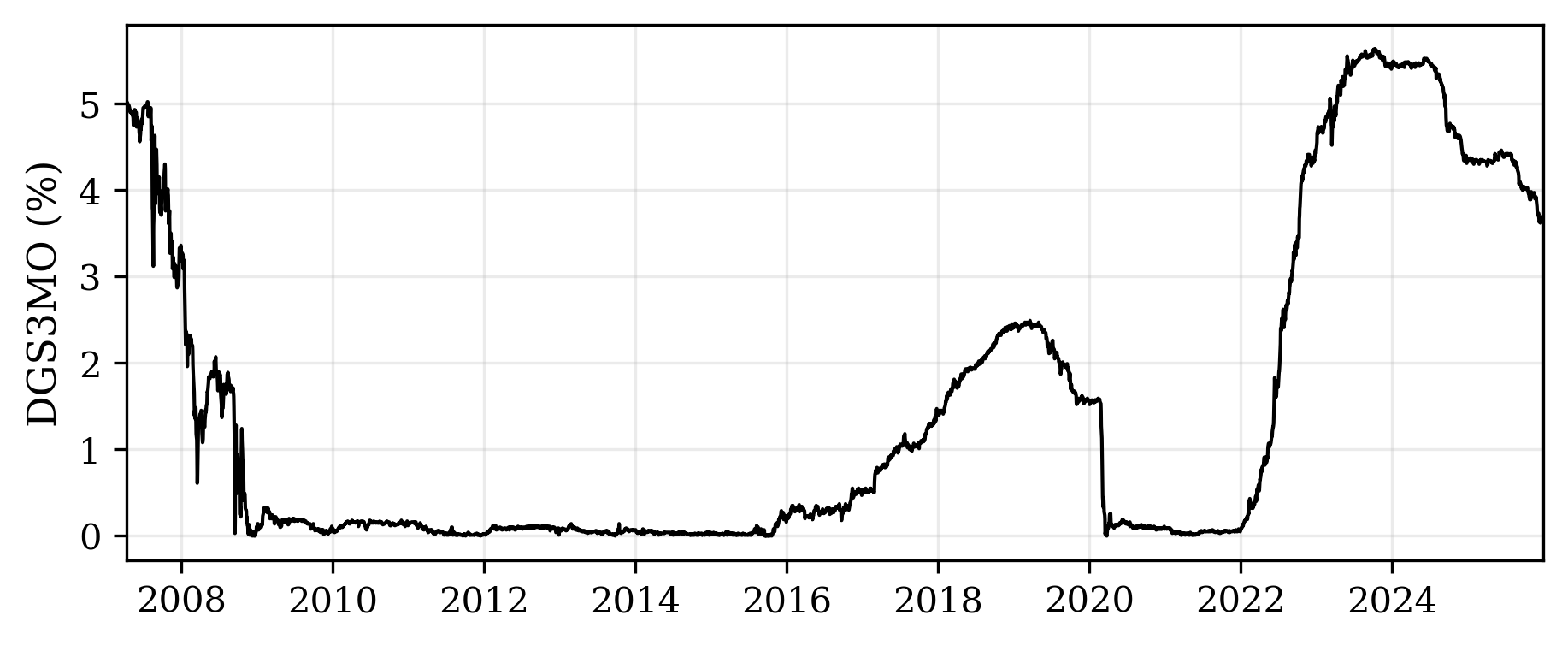}
	\caption{Evolution of the 3-month U.S. Treasury constant maturity rate}
	\label{fig:rf}
\end{figure}

\FloatBarrier

\subsection{Rolling Execution and Transaction Costs}\label{sec:execution}

The out-of-sample period contains 3,955 earned returns from April 14, 2010 to December 31, 2025 and 198 scheduled decisions. Let $\boldsymbol r(t)$ denote the vector of simple asset returns on date $t$, with $n=12$ components. The historical window has length $\ell=756$ and targets are updated every 20 trading days. Let $t$ denote the first return date of a newly formed portfolio. Its information set is
\begin{equation*}
\mathcal{D}_t=\{\boldsymbol{r}(t-\ell-1),\ldots,\boldsymbol{r}(t-2)\}.
\end{equation*}
The information set ends at the close of date $t-2$. Targets are computed without using the return on date $t-1$, executed at the close of $t-1$, and first earn the return on date $t$. Neither the execution-interval return nor the first earned return enters training. Figure~\ref{fig:rf} shows the Treasury-yield proxy over the sample: extended near-zero periods contrast with the rise above 5\% in 2023--2024. The opportunity-cost benchmark therefore varies materially over time rather than remaining a fixed zero-return threshold. Writing the annual percentage yield as $y(t)$, the daily benchmark approximation uses its lagged value:
\begin{equation*}
r_f(t)=\left(1+\frac{y(t-2)}{100}\right)^{1/252}-1.
\end{equation*}
The dated yield is assumed available by execution. This exponential conversion of an investment-basis quote is a benchmark approximation, not an observed Treasury-bill holding-period return. The same lagged daily value is subtracted from every scenario at a decision. A fixed-weight sensitivity check compares it with the annual percentage yield divided by 100 and 252; this check does not include portfolio reoptimization.

Targets are long-only and fully invested. The weight vector $\boldsymbol w^{\mathrm{held}}(s)$ that earns the return on date $s$ is allowed to drift between decisions. Gross portfolio return and post-return component weights are
\begin{align*}
r_p(s)&=\boldsymbol w^{\mathrm{held}}(s)^\top\boldsymbol r(s), \\
w_i(s^+)&=\frac{w_i^{\mathrm{held}}(s)[1+r_i(s)]}{1+r_p(s)}.
\end{align*}

At rebalancing, the carried pre-trade vector $\boldsymbol w^{-}(t)$ is traded to the new target $\boldsymbol w^{\mathrm{tar}}(t)$. The reporting index below runs over the realized sample, and its length is denoted by $T$. One-way turnover (TO) and the transaction-cost fraction (TC) are
\begin{align*}
\mathrm{TO}(t)&=\frac12\sum_{i=1}^{n}|w_i^{\mathrm{tar}}(t)-w_i^{-}(t)|, \\
\mathrm{TC}(t)&=2c\,\mathrm{TO}(t),\qquad
\mathrm{TO}_{\mathrm{total}}=\sum_{t=1}^{T}\mathrm{TO}(t),
\end{align*}
where $c=0.001$ is the cost rate of 10 basis points per unit of total purchases plus sales. Initial formation is not charged, turnover is zero outside rebalancing dates, and the fee is attributed to the first subsequent earned return. Net return and normalized wealth follow
\begin{equation}
\begin{aligned}
r_p^{\mathrm{net}}(t)&=[1-\mathrm{TC}(t)][1+r_p(t)]-1, \\
P_p(t)&=\prod_{s=1}^{t}[1+r_p^{\mathrm{net}}(s)],\qquad P_p(0)=1.
\end{aligned}
\label{eq:net_wealth}
\end{equation}
The net-return and wealth recursions in Equation~\eqref{eq:net_wealth} use a small-cost approximation: exact self-financing trades after a fee deduction can differ slightly from pre-cost weight differences. For example, reallocating $(0.6,0.4)$ to $(0.4,0.6)$ produces one-way turnover 0.2, total traded value 0.4 and a cost fraction 0.0004. The reported first-day net return for a 1\% gross return is
\begin{equation*}
(1-0.0004)(1.01)-1=0.9596\%.
\end{equation*}
Costs are evaluated after weight selection; they are not an allocation penalty. Realized performance measures are defined in Section~\ref{sec:performance_measures}.

\subsection{Data Provenance and Computational Reproducibility}\label{sec:reproducibility}

The dataset was downloaded on October 8, 2026 using yfinance 1.7.0. Training used PyTorch 2.11.0 on an NVIDIA Tesla T4 graphics processing unit (GPU), with about 2.62 hours of cumulative model fitting. Portfolio optimization used SciPy 1.16.3 on the central processing unit (CPU).

The accompanying code and \texttt{REPRODUCIBILITY.txt} document preprocessing, training, scenario construction, optimization, execution, and metric calculation. They identify the fixed data snapshot, software environment, random seeds and saved inputs, targets and solver records needed to reproduce selected windows.

\section{Portfolio Criteria and Scenario Distributions}\label{sec:methods}

\subsection{Constraints, Tail Measures, and Benchmarks}\label{sec:prelim}

For the asset-return vectors introduced in Section~\ref{sec:execution}, the feasible set of portfolio weights is
\begin{equation}
\mathcal{W}=\left\{\boldsymbol{w}\in\mathbb{R}^{n}:\sum_{i=1}^{n}w_i=1,\quad 0\leq w_i\leq1\right\}.
\label{eq:constraints}
\end{equation}
The constraints in Equation~\eqref{eq:constraints} exclude short sales and borrowing. The risk-free proxy is a benchmark, not an investable thirteenth asset. Let $R_p$ be a candidate portfolio's random return, $F$ its cumulative distribution function and $\alpha\in(0,1)$ a lower-tail probability. The loss value-at-risk (VaR) and its quantile-integral CVaR are
\begin{align*}
\textrm{VaR}_{\alpha}(R_p) &= - \inf\{x\in\mathbb{R}\mid F(x)\ge\alpha\}, \\
\textrm{CVaR}_{\alpha}(R_p) &= \frac{1}{\alpha}\int_{0}^{\alpha}\textrm{VaR}_{\gamma}(R_p)\textrm{ d}\gamma.
\end{align*}
These definitions apply to empirical distributions through fractional cutoff weighting~\citep{rockafellar2000optimization}, specified in Section~\ref{sec:rrp}. The daily Sharpe ratio (SR; \citealp{sharpe1966mutual}) and Sortino ratio (ST; \citealp{sortino2001managing}) are
\begin{align*}
\textrm{SR}(T) &= \frac{\mathbb{E}\left[r_p(t) - r_f(t)\right]}{\sqrt{\textrm{Var}\left[r_p(t) - r_f(t)\right]}} = \frac{\mu_p}{\sigma_p}, \\
\textrm{ST}(T)
&=
\frac{
\mathbb{E}\left[r_p(t)-r_f(t)\right]
}{
\sqrt{
\mathbb{E}\left[
\min\left(r_p(t)-r_f(t),0\right)^2
\right]
}
},
\end{align*}
where $T$ is the observation count, and $\mu_p$ and $\sigma_p$ are the mean and standard deviation of excess returns. Empirical evaluation replaces expectations and variances with sample estimates, using one fewer degree of freedom for variances. Optimization uses scenario returns at fixed weights and the common window-specific risk-free input; Section~\ref{sec:performance_measures} gives the annualized sample formulas for realized net returns.

The equally weighted portfolio (EWP) resets to equal weights at each scheduled rebalance; the equal-weight buy-and-hold portfolio (EWBH) forms equal weights once and subsequently drifts:
\begin{align*}
w_{i}^{\mathrm{EWP}}&=\frac{1}{n},
\qquad i=1,\ldots,n, \\
w_{i}^{\mathrm{EWBH}}(0)&=\frac{1}{n},
\qquad i=1,\ldots,n.
\end{align*}

The historical tail-ratio allocation is the direct optimized comparator for generated policies. EWP and EWBH provide simple investable references under the common execution and cost rules. Section~\ref{sec:resampling} separately compares scenario sources while holding the tail-ratio objective fixed.

\subsection{Modified Rachev Ratio and Scenario Weighting}\label{sec:rrp}

The theoretical Rachev criterion compares expected favorable and unfavorable extreme excess returns~\citep{biglova2004different,rachev2008advanced}. Let $\alpha$ and $\beta$ denote the favorable- and unfavorable-tail probabilities, respectively. Using the loss-CVaR sign convention, its components are expected tail loss (ETL) and expected tail reward (ETR):
\begin{align*}
\mathrm{ETL}_{\beta}(\boldsymbol{w})
&=
\mathrm{CVaR}_{\beta}\left(\boldsymbol{w}^{\top}\boldsymbol{r}-r_f\right), \\
\mathrm{ETR}_{\alpha}(\boldsymbol{w})
&=
\mathrm{CVaR}_{\alpha}\left(r_f-\boldsymbol{w}^{\top}\boldsymbol{r}\right).
\end{align*}
Equation~\eqref{eq:theoretical_ratio} defines the theoretical ratio and its formal allocation objective:
\begin{equation}
\begin{aligned}
\mathrm{RR}_{\alpha,\beta}(\boldsymbol{w})
&=
\frac{
\mathrm{ETR}_{\alpha}(\boldsymbol{w})
}{
\mathrm{ETL}_{\beta}(\boldsymbol{w})
}
=
\frac{
\mathrm{CVaR}_{\alpha}\left(r_f-\boldsymbol{w}^{\top}\boldsymbol{r}\right)
}{
\mathrm{CVaR}_{\beta}\left(\boldsymbol{w}^{\top}\boldsymbol{r}-r_f\right)
}, \\
\boldsymbol{w}^{\mathrm{RRP}}
&=
\arg\max_{\boldsymbol{w}\in\mathcal{W}}
\quad
\mathrm{RR}_{\alpha,\beta}(\boldsymbol{w}).
\end{aligned}
\label{eq:theoretical_ratio}
\end{equation}

\begin{figure} 
\centering
\includegraphics[width=.73\textwidth]{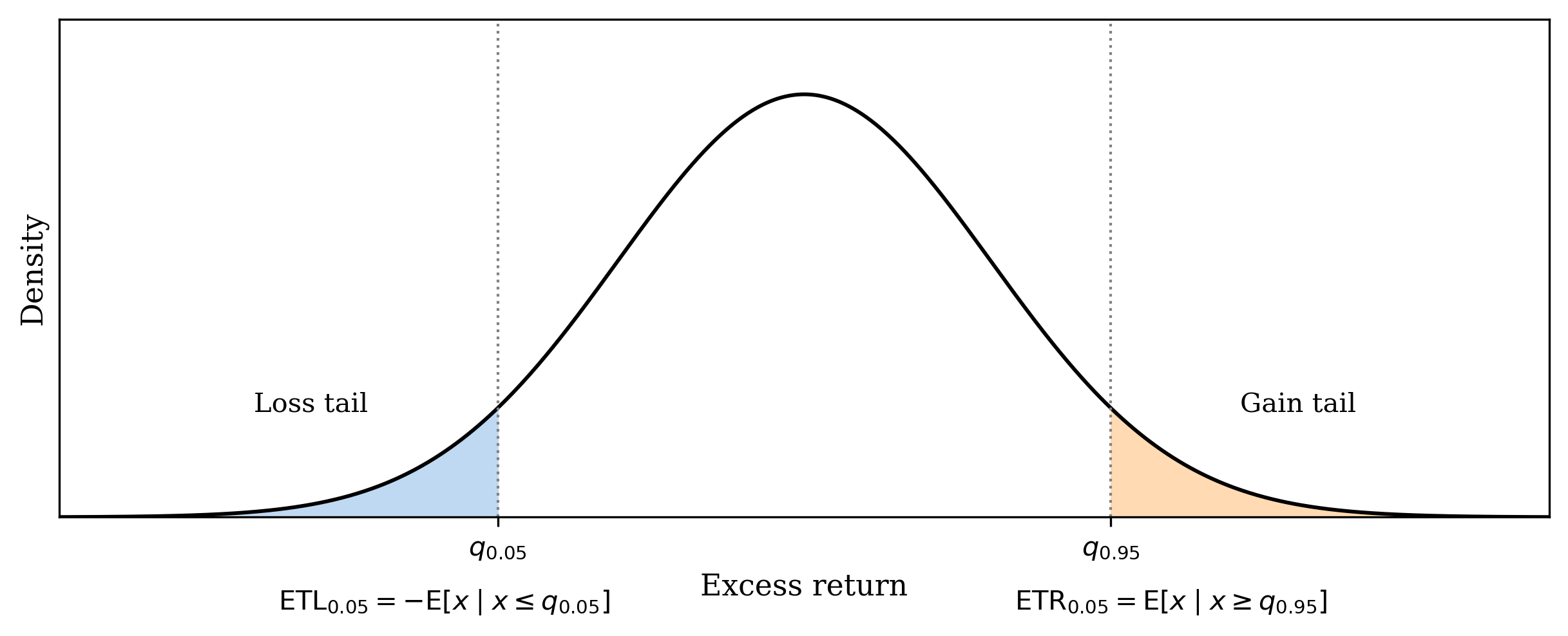}
\caption{Schematic illustration of the gain and loss tails used in the Rachev ratio. The density is a continuous-distribution illustration; empirical cutoff ties require appropriate probability weighting and do not impose normality}
\label{fig:rachev_ratio_illustration}
\end{figure}

Figure~\ref{fig:rachev_ratio_illustration} illustrates favorable and unfavorable tails for a continuous distribution. The left and right shaded regions in Figure~\ref{fig:rachev_ratio_illustration} identify the unfavorable and favorable outcomes whose conditional means enter the loss and reward components, respectively. The figure does not assume Gaussian returns; empirical cutoff ties require the fractional probability weights defined below.

The loss denominator is economically meaningful when positive. The empirical procedure below additionally specifies an absolute value and a numerical floor; outside the positive-loss domain it is a modified criterion, not the theoretical ratio.

Let $M$ be the generated count and $\eta\in[0,1]$ the synthetic probability mass, varied independently. Write $r_j$ and $\widetilde r_j$ for historical and generated asset-return vectors, respectively, and $\delta$ for a unit point mass. TimeGAN is fitted to the available historical window to produce the synthetic sample; its architecture and training are described in Section~\ref{sec:timegan_framework}. The resulting scenario distribution is
\begin{equation}
F_{\eta,M}=(1-\eta)\frac1\ell\sum_{j=1}^{\ell}\delta_{r_j}
           +\eta\frac1M\sum_{j=1}^{M}\delta_{\widetilde r_j}.
\label{eq:scenario_distribution}
\end{equation}
In Equation~\eqref{eq:scenario_distribution}, historical-only allocation sets synthetic mass to zero; replacement uses only generated rows; hybrids retain both sources. Rows within each source have equal probability, and the two sources receive their specified total masses. Increasing draw count at fixed mass changes Monte Carlo resolution, not the weight of the generator or the amount of independent market information.

Let $\boldsymbol r_j$ now range over the scenario vectors in this distribution, with normalized row probabilities $p_j$. Set the common empirical tail probability to 5\%, and let $c_L$ be the first weighted order statistic attaining that lower-tail mass. The scenario excess returns and fractional lower-tail coefficients are
\begin{align*}
x_j(\boldsymbol w)&=\boldsymbol w^\top\boldsymbol r_j-r_f,\qquad \alpha=0.05, \\
q_j^L&=\frac{p_j}{\alpha}\left[\mathbf{1}\{x_j<c_L\}+\kappa_L\mathbf{1}\{x_j=c_L\}\right], \\
\kappa_L&=\frac{\alpha-\sum_{x_j<c_L}p_j}{\sum_{x_j=c_L}p_j},\qquad
\sum_jq_j^L=1,
\end{align*}
where $\kappa_L$ is the fraction of probability retained at the cutoff. Upper coefficients $q_j^U$ are defined analogously with the inequalities reversed. Cutoff ties receive probability in proportion to their scenario weights. The empirical means and implemented objective are
\begin{equation}
\begin{aligned}
\bar x_L(\boldsymbol w)&=\sum_jq_j^Lx_j(\boldsymbol w),\qquad
\bar x_U(\boldsymbol w)=\sum_jq_j^Ux_j(\boldsymbol w), \\
\widehat{\mathrm{RR}}(\boldsymbol w;F_{\eta,M})
&=\frac{\bar x_U(\boldsymbol w)}{\max\{|\bar x_L(\boldsymbol w)|,\varepsilon\}},
\qquad \varepsilon=10^{-8}.
\end{aligned}
\label{eq:weighted_rr}
\end{equation}
The same fractional convention is used for realized tail ratios. Portfolio targets under replacement and supplementation remain explicitly defined by
\begin{align*}
\boldsymbol{w}^{\mathrm{AI\mbox{-}RRP}}(t)&=\arg\max_{\boldsymbol{w}\in\mathcal{W}}\widehat{\mathrm{RR}}(\boldsymbol{w};\widetilde{\mathcal{D}}_t), \\
\boldsymbol{w}^{\mathrm{Hybrid\mbox{-}RRP}}(t)&=\arg\max_{\boldsymbol{w}\in\mathcal{W}}\widehat{\mathrm{RR}}(\boldsymbol{w};\mathcal{H}_t),
\end{align*}
where $\widetilde{\mathcal D}_t$ is the generated empirical distribution ($\eta=1$) and $\mathcal H_t$ is the historical--synthetic distribution with the specified hybrid mass. The numerical procedure in Section~\ref{sec:numerical_search} selects feasible candidates for these objectives; the argmax notation is not a claim that a global maximum has been attained.

\begin{figure}[H]
\centering
\includegraphics[width=.84\textwidth]{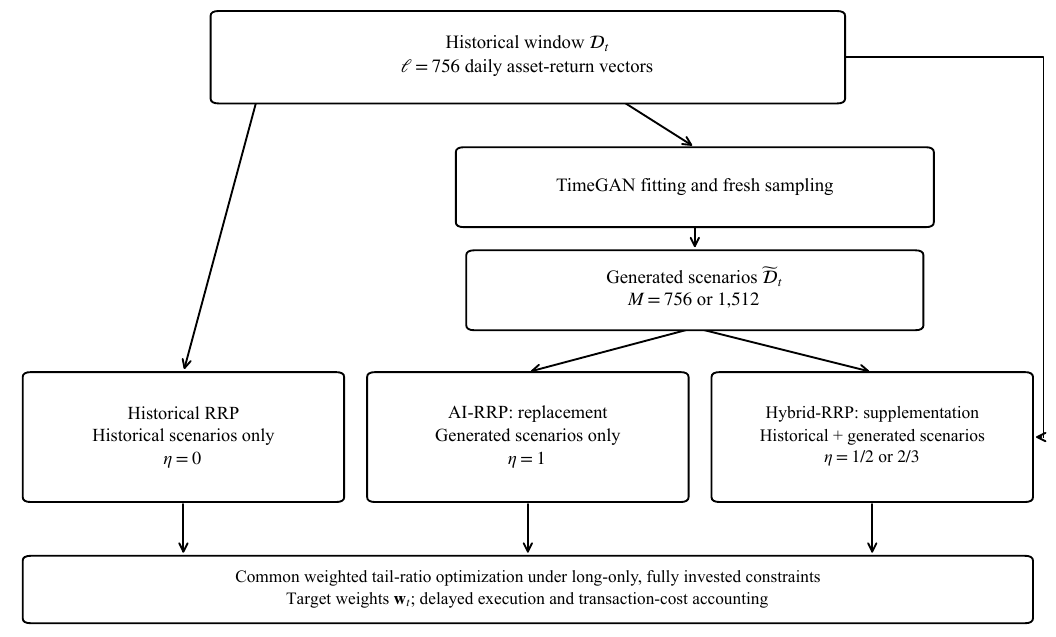}
\caption{Construction of historical RRP, AI-RRP and Hybrid-RRP in the completed experiment. Arrows show the sources of scenarios supplied to the common allocation procedure. The synthetic probability mass $\eta$ and generated count $M$ are specified separately; within each source, observations receive equal weights.}
\label{fig:ai_hybrid_rrp_construction}
\end{figure}

Figure~\ref{fig:ai_hybrid_rrp_construction} summarizes the three construction paths. For both generated designs, TimeGAN is fitted to the same available historical window. Historical RRP uses only historical observations, AI-RRP uses only generated observations in optimization, and Hybrid-RRP retains both sources with total probability masses $1-\eta$ and $\eta$, respectively. In the completed experiment, $M\in\{756,1512\}$ varies independently of the hybrid mass $\eta\in\{1/2,2/3\}$. The weighting is applied to scenarios before optimization; it does not average separately optimized portfolio weights.

\subsection{Realized Performance Measures}\label{sec:performance_measures}

Realized performance uses the $T$ net returns and wealth path from Section~\ref{sec:execution}, including costs and intervening weight drift. With $\bar r_p^{\mathrm{net}}$ denoting the sample mean, annualized geometric net return and sample volatility are
\begin{align*}
r_p^{\mathrm{ann}}&=P_p(T)^{252/T}-1, \\
\sigma_p^{\mathrm{ann}}&=\sqrt{252}\left[\frac{1}{T-1}\sum_{t=1}^{T}(r_p^{\mathrm{net}}(t)-\bar{r}_p^{\mathrm{net}})^2\right]^{1/2}.
\end{align*}

Applying the SR and ST definitions to daily net excess returns gives their annualized sample forms:
\begin{align*}
e_t&=r_p^{\mathrm{net}}(t)-r_f(t),\qquad \bar e=T^{-1}\sum_{t=1}^{T}e_t, \\
\mathrm{SR}^{\mathrm{ann}}&=\sqrt{252}\,\frac{\bar{e}}{s_e},\qquad
s_e=\left[\frac{1}{T-1}\sum_{t=1}^{T}(e_t-\bar{e})^2\right]^{1/2}, \\
\mathrm{ST}^{\mathrm{ann}}&=\sqrt{252}\,\frac{\bar{e}}{\left[T^{-1}\sum_{t=1}^{T}\min(e_t,0)^2\right]^{1/2}}.
\end{align*}
A floor of $10^{-12}$ is applied to both denominators. The daily target is the lagged risk-free rate and downside deviation includes all observations. The realized RR applies the weighted tail means and ratio in Equation~\eqref{eq:weighted_rr} to the net excess returns $e_t$, assigning probability $1/T$ to each date and retaining the same $10^{-8}$ floor. Maximum drawdown (MDD) and daily loss-CVaR are reported as losses, not annualized:
\begin{align*}
\mathrm{MDD}&=\max_{1\leq t\leq T}\left\{1-\frac{P_p(t)}{\max_{0\leq s\leq t}P_p(s)}\right\}, \\
\widehat{\mathrm{CVaR}}_{0.05}^{\mathrm{daily}}
&=-\sum_{j=1}^{T}q_j^{L}\,r_p^{\mathrm{net}}(j),
\end{align*}
where the lower-tail coefficients defined in Section~\ref{sec:rrp} are recomputed from net returns themselves with equal date probabilities, without subtracting the risk-free rate. The same fractional convention is used throughout portfolio evaluation.

\section{TimeGAN Construction and Scenario Fidelity}\label{sec:generation}

\subsection{TimeGAN Architecture and Training}\label{sec:timegan_framework}

The fitted architecture is an instrumented PyTorch TimeGAN variant, not an exact reproduction of the original TensorFlow implementation~\citep{yoon2019time}. Table~\ref{tab:timegan_parameters} summarizes its settings. It contains an embedder, recovery network, noise generator, temporal supervisor, and discriminator. Single-layer gated recurrent units (GRUs) have linear projection outputs; recovery is unbounded in standardized return units and discriminator projections are logits. Independent standard-normal noise is used. Each historical window is standardized with its own per-asset mean and population standard deviation, and produces 737 overlapping sequences of length 20.

\begin{table}[htbp]
\centering
\small
\caption{TimeGAN model and training parameters}
\label{tab:timegan_parameters}

\begin{tabular}{ll}
\toprule
Parameter & Value \\
\midrule
Sequence length & 20 trading days \\
Number of GRU layers & 1 \\
Hidden dimension & 32 \\
Noise dimension & 8 \\
Noise distribution & Independent standard normal \\
GRU projection outputs & Linear; discriminator logits \\
\midrule
Batch size & 128 \\
Training epochs & 50 per training stage \\
Optimizer & Adaptive moment estimation (Adam) \\
Embedder--recovery learning rate & \(10^{-4}\) \\
Generator--supervisor learning rate & \(10^{-4}\) \\
Discriminator learning rate & \(10^{-4}\) \\
\midrule
Adversarial-loss weight \(\gamma\) & 1 \\
Moment-loss weight & 100 \\
Supervised-loss weight & 100 \\
\midrule
Base training seeds & 12345, 23456, 34567, 45678, 56789 \\
\bottomrule
\end{tabular}
\end{table}

Each of the embedding, supervised and joint stages runs for 50 epochs. The final incomplete minibatch is retained, giving six minibatches per epoch and 300 updates per applicable component/stage. The losses and update rules are specified below. The five base seeds and all hyperparameters are fixed across the design; primary seed 12345 is used for diagnostic illustrations, not chosen by performance.

The training configuration is held fixed across rolling windows and allocation policies to maintain a consistent experimental protocol, rather than retuned to obtain the most favorable downstream outcome in each window. This makes the scenario-use comparisons controlled conditional on the chosen implementation; it does not establish convergence, sufficient training or optimal model adequacy. Implementation correctness and suitability for this financial task are distinct questions. The architecture, losses and diagnostic records make the protocol inspectable, while scenario fidelity and decision consequences must be evaluated separately.

Each completed fit produces three draws with separately recorded sampling seeds. A draw contains 1,512 daily asset vectors from 76 independent 20-day noise sequences, concatenated and truncated. The 756-row design uses the nested prefix of the same draw; optimization uses replicate zero, while all three draws enter quality diagnostics. Normalization is reversed and a six-standard-deviation clipping safeguard is applied. Asset vectors remain intact; no continuity is assumed across independent sequence boundaries. The single-period allocation objective uses the joint distribution of daily vectors, so serial fidelity is diagnostic rather than a direct objective input. More simulated rows do not supply additional independent historical tail events.

Write the embedder, recovery network, generator, supervisor, and discriminator as $E,R,G,S,D$. Let $X$ be a standardized historical return sequence of length $L=20$ and $Z$ an independent standard-normal noise sequence. The embedded historical sequence $H$, generated latent sequence $\widehat E$, supervised latent sequence $\widehat H$ and recovered synthetic return sequence $\widehat X$ are
\begin{align*}
H&=E(X),&\widehat E&=G(Z),&\widehat H&=S(\widehat E),&\widehat X&=R(\widehat H).
\end{align*}
Using mean squared error (MSE), the supervised loss is
\begin{equation*}
L_S=\operatorname{MSE}(H_{2:L},S(H)_{1:L-1}).
\end{equation*}

The moment discrepancy $L_V$ sums mean absolute differences in asset means and population standard deviations over minibatch and sequence axes. Let $B(\cdot,b)$ denote binary cross entropy on discriminator logits with target label $b\in\{0,1\}$. The implemented generator, recovery and discriminator losses are
\begin{equation}
\begin{aligned}
L_G&=B(D(\widehat H),1)+\gamma B(D(\widehat E),1)
       +100\sqrt{L_S+10^{-8}}+100L_V, \\
L_{ER}&=\sqrt{\operatorname{MSE}(R(E(X)),X)+10^{-8}}+0.1L_S, \\
L_D&=B(D(H),1)+B(D(\widehat H),0)+\gamma B(D(\widehat E),0),\quad\gamma=1.
\end{aligned}
\label{eq:training_losses}
\end{equation}
The loss functions in Equation~\eqref{eq:training_losses} distinguish adversarial, supervised, and moment-matching components. Embedding pretraining uses the reconstruction square-root term alone, and supervised pretraining updates the supervisor using its supervised loss. Separate Adam optimizers have learning rate $10^{-4}$. Joint training performs one generator--supervisor, embedder--recovery, and discriminator update per retained minibatch. Recovery projections are linear and therefore compatible with positive and negative standardized returns; the bounded-support sigmoid concern does not describe this implementation. Normal noise and per-window standardization differ from the original TensorFlow min--max implementation~\citep{yoon2019time}.

All current policy comparisons retain the final incomplete minibatch. Earlier fits with a different update count are excluded from the matched comparisons.

\subsection{Marginal, Cross-Asset, and Temporal Diagnostics}\label{sec:quality}

Diagnostics average over windows, training seeds and three independent draws at two generated counts. Overlapping historical windows make these records dependent. Let $\phi_i$ be an asset-level statistic, $\rho_{ij}$ a pairwise return correlation and $b$ a generated-draw index. The corresponding mean absolute error (MAE) measures are
\begin{align*}
\mathrm{MAE}_{\phi}(t,b)&=\frac{1}{n}\sum_{i=1}^{n}
\left|\phi_i(\widetilde{\mathcal{D}}_{t,b})-\phi_i(\mathcal{D}_t)\right|, \\
\mathrm{MAE}_{\mathrm{corr}}(t,b)&=\frac{1}{n(n-1)}
\sum_{i=1}^{n}\sum_{\substack{j=1\\j\ne i}}^{n}
\left|\rho_{ij}(\widetilde{\mathcal{D}}_{t,b})-\rho_{ij}(\mathcal{D}_t)\right|.
\end{align*}
Signed errors use generated minus historical values. Return-distribution errors are expressed in daily percentage points (pp/day). Positive tail-loss error means overstatement of losses. Marginal volatility uses population standard deviations, consistent with training normalization; realized portfolio volatility uses the sample convention in Section~\ref{sec:performance_measures}.

\begin{figure}[!htbp]
\centering
\includegraphics[width=.90\textwidth,height=.58\textheight,keepaspectratio]{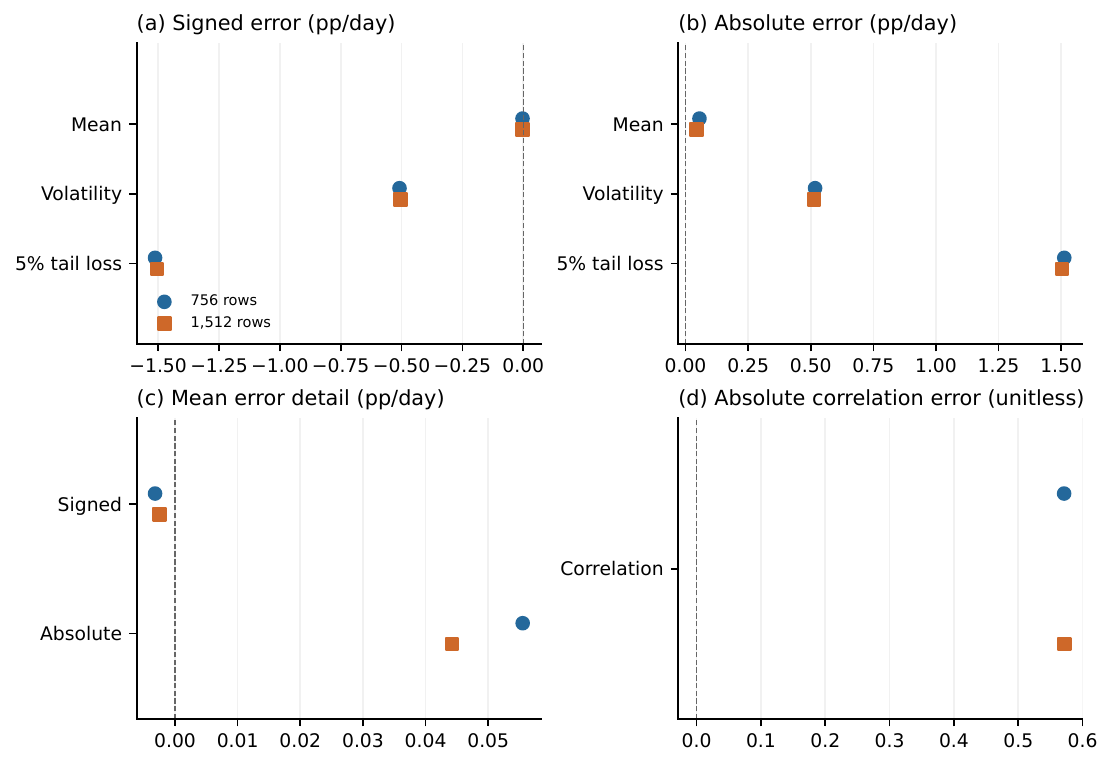}
\caption{Scenario fidelity at 756 and 1,512 generated rows. Panels (a)--(c) use daily-return percentage points; panel (d) is unitless. Each point averages 2,970 window--seed--draw records. Signed errors are generated minus historical. Marginal descriptive tail errors use integer-ceiling 5\% tails, unlike the fractional allocation and realized-ratio convention. Panel (c) enlarges the mean-error scale; axes are not shared across panels.}\label{fig:fidelity_summary}
\end{figure}

Figure~\ref{fig:fidelity_summary} shows that marginal means are close in aggregate, but volatility is understated by about 0.50 daily percentage points and 5\% marginal tail losses by about 1.50 points. Mean absolute correlation error is approximately 0.571 at both counts. Increasing draw count therefore does not repair the fitted dependence structure or the systematic tail discrepancy.

\begin{figure}[!htbp]\centering
\includegraphics[width=.88\textwidth,height=.58\textheight,keepaspectratio]{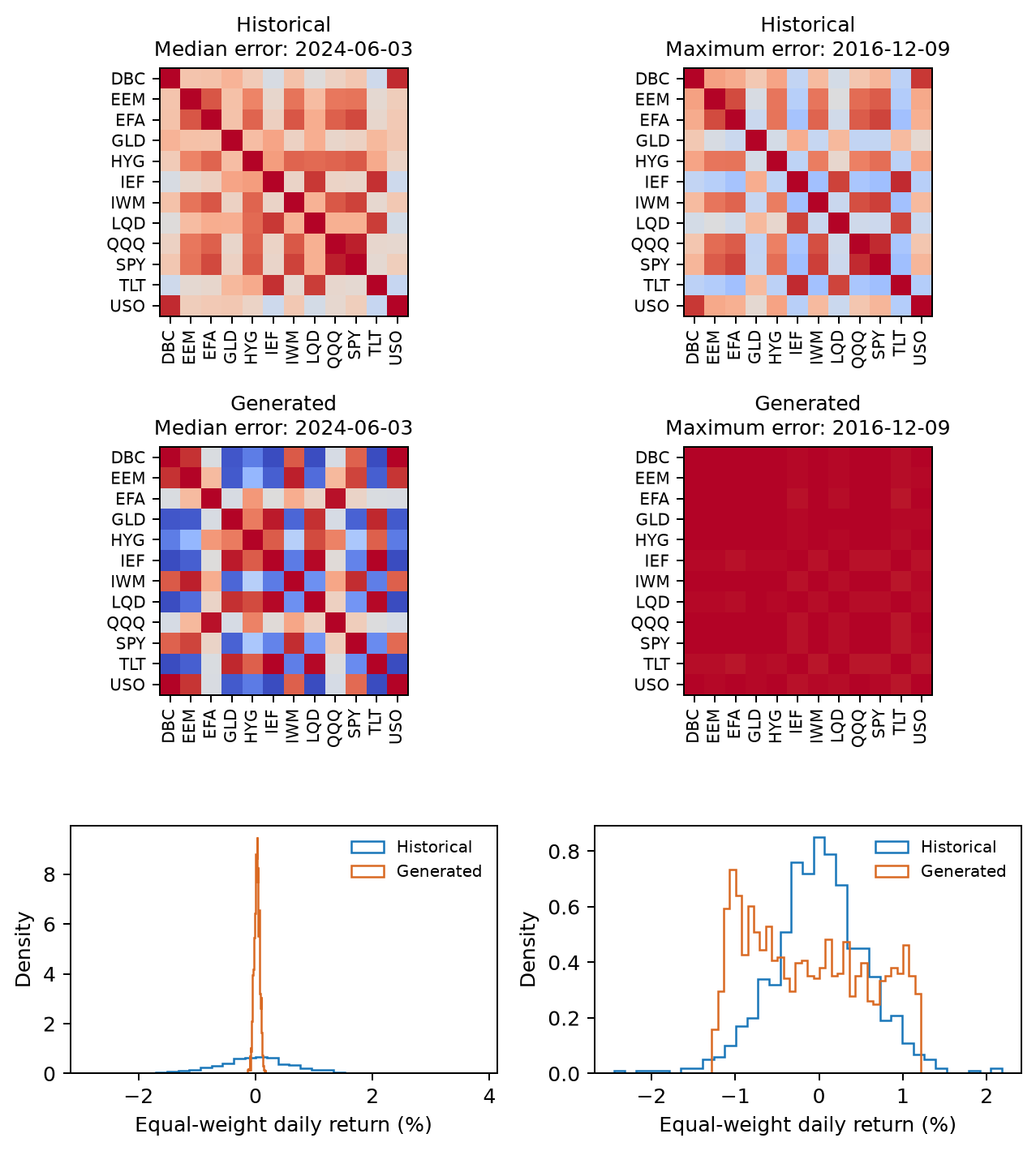}
\caption{Historical and generated dependence and equal-weight daily return distributions at the median- and maximum-correlation-error windows, seed 12345. Selection uses all three 756-row replicates; plotted raw samples use replicate zero and 1,512 rows.}\label{fig:s_dependence}
\end{figure}

\begin{figure}[!htbp]\centering
\includegraphics[width=.88\textwidth,height=.58\textheight,keepaspectratio]{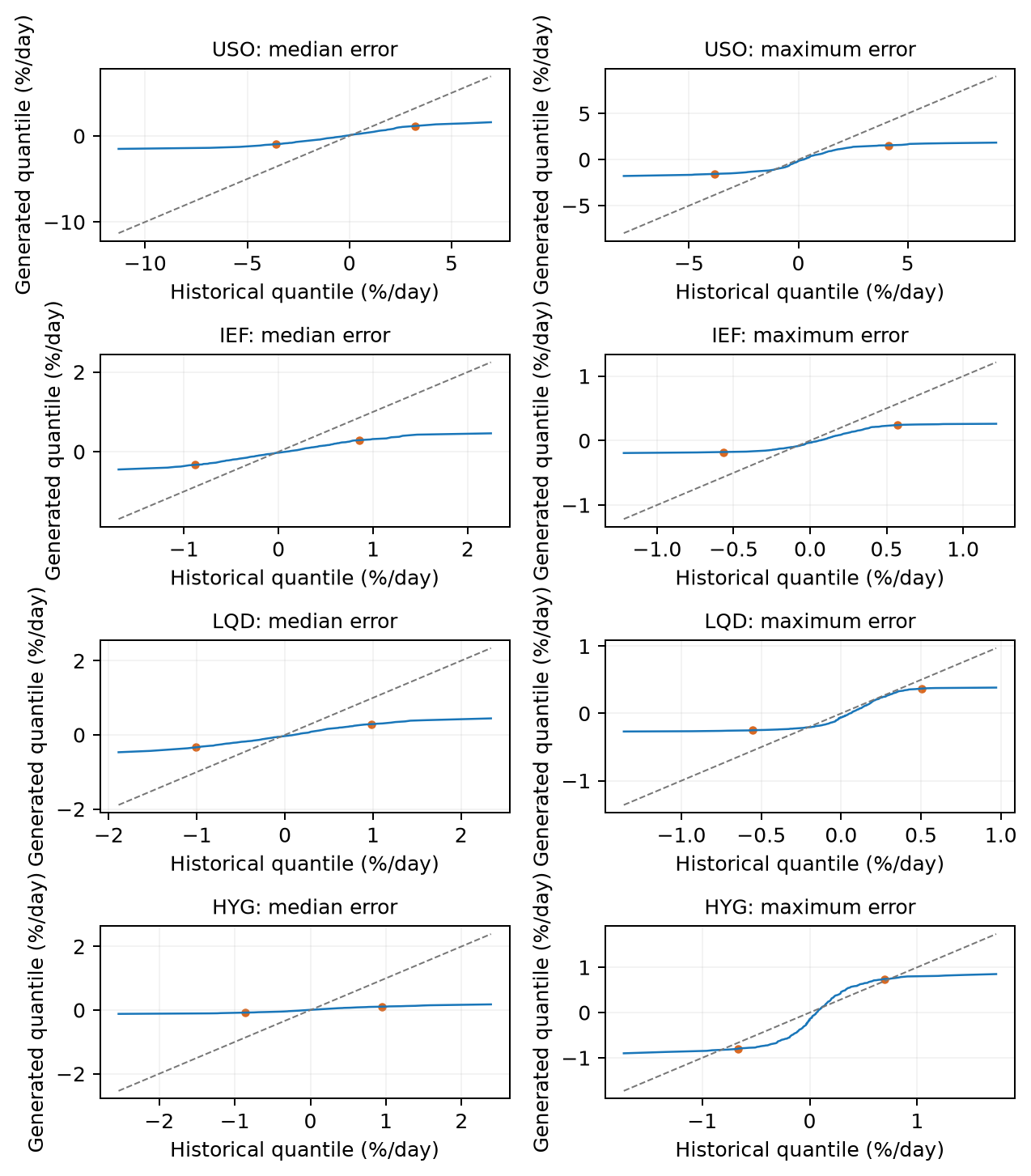}
\caption{Marginal quantile comparisons for USO, IEF, LQD and HYG at the same selected windows. Both axes are daily-return percentages; the diagonal is equality and dots mark 5\% and 95\% quantiles.}\label{fig:s_marginals}
\end{figure}

Representative windows are selected separately for each seed by median and maximum mean correlation MAE over the three 756-row replicates, without using investment performance. Figure~\ref{fig:s_dependence} contrasts historical and generated correlation matrices at the primary seed's selected windows using the first 1,512-row replicate. At the median-error window, several generated correlations have signs unlike their historical counterparts, and the equal-weight generated-return histogram is sharply compressed. At the maximum-error window, generated correlations are almost uniformly strongly positive, erasing much of the historical cross-asset structure. Figure~\ref{fig:s_marginals} complements this joint-distribution view: the USO, IEF, LQD and HYG quantile curves flatten in the tails relative to the equality line, showing compression of extreme returns. These examples illustrate different forms of distortion; they do not establish the shape of every marginal distribution.

\begin{figure}[!htbp]
\centering
\includegraphics[width=.88\textwidth,height=.58\textheight,keepaspectratio]{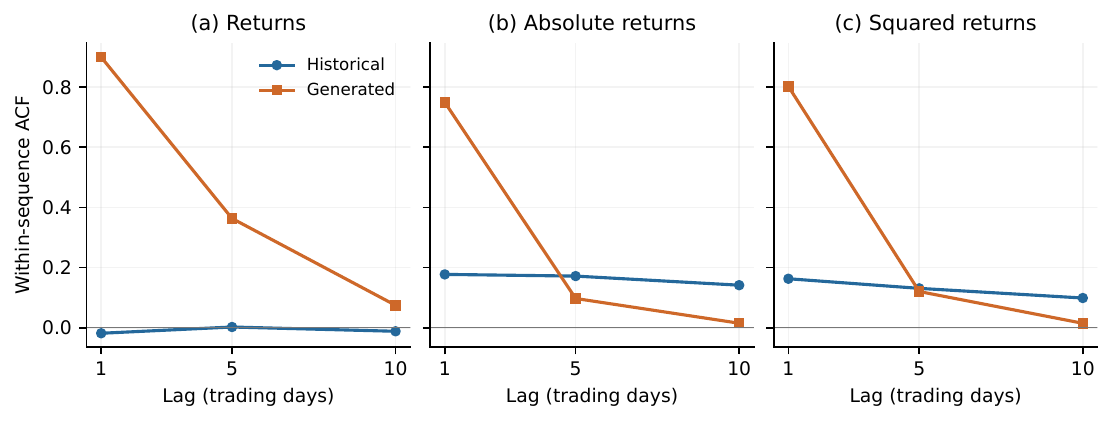}
\caption{Within-sequence autocorrelations for returns, absolute returns and squared returns at lags 1, 5 and 10. Values average assets and 30 draws from ten selected representative models, not all 990 fits. Historical pairs use contiguous dates; generated pairs crossing independent sequence boundaries are excluded. Lines connect the reported lags.}\label{fig:temporal_fidelity}
\end{figure}

Within-sequence autocorrelation function (ACF) estimates summarize dependence in returns and their absolute and squared transformations. Figure~\ref{fig:temporal_fidelity} shows marked temporal discrepancies in the selected representative models: mean first-lag return ACF is 0.900 for generated sequences and -0.018 for historical observations. Absolute and squared generated returns also have much stronger first-lag dependence, but their ACFs fall below historical estimates at lag 10. Thus, these examples reveal distortion of persistence as well as cross-asset dependence.

\FloatBarrier

\subsection{Training Effort and Finite-Generation Variability}\label{sec:training_variation}
\begin{figure}[!htbp]
\centering
\includegraphics[width=.84\textwidth]{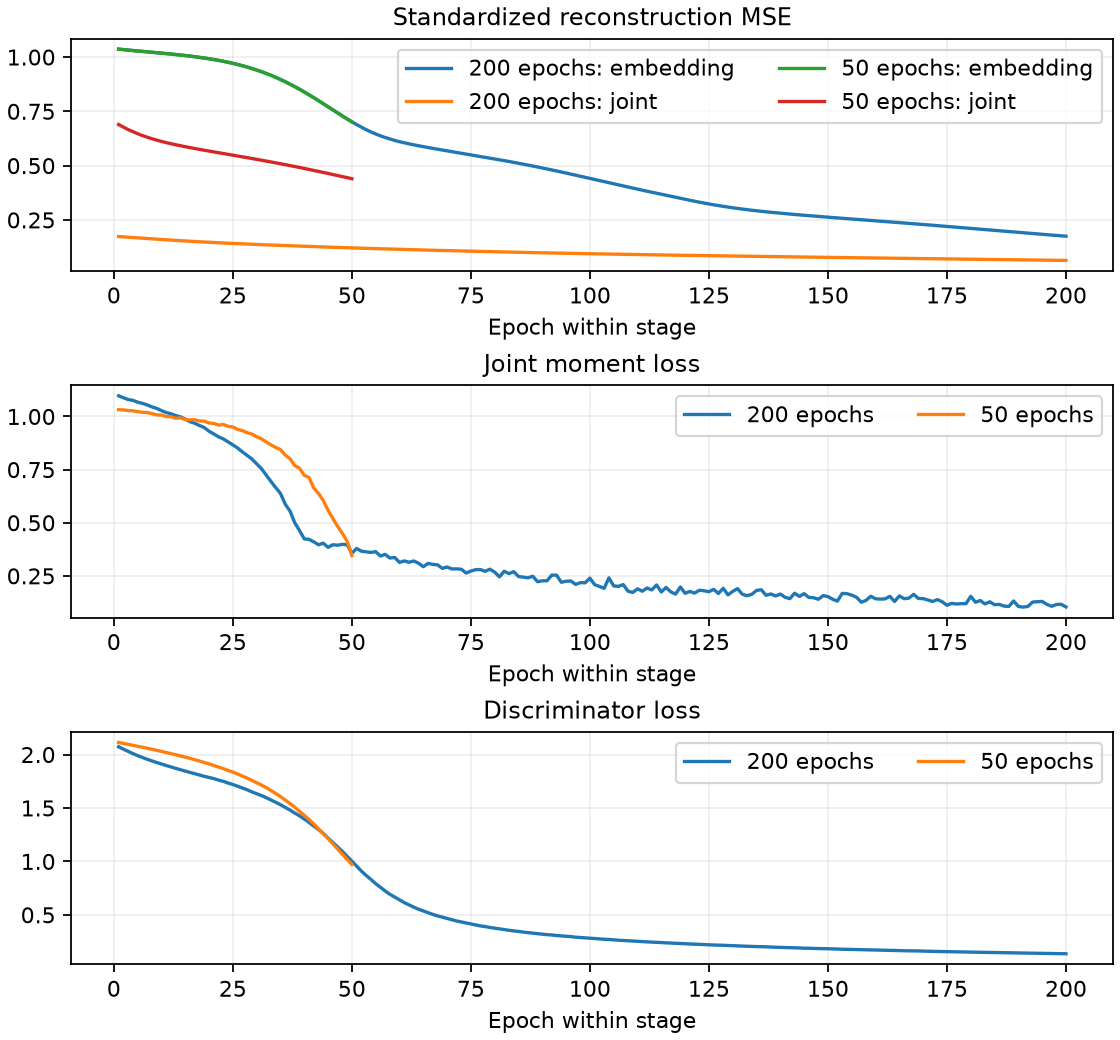}
\caption{Instrumented reconstruction and joint-training losses in the first chronological window, seed 12345. Epochs are counted separately within each stage. These losses do not certify distributional fidelity.}
\label{fig:review_training}
\end{figure}

The limited longer-training diagnostics below were conducted before the complete GPU experiment. They evaluate fitting and sampling behavior, not full-period portfolio performance. Appendix Figure~\ref{fig:supp_training_pairs} shows that raising training effort from 50 to 200 epochs in two additional primary-seed windows lowered reconstruction MSE from 0.293 to 0.039 and from 0.319 to 0.031. Correlation MAE nevertheless increased from 0.775 to 0.778 and from 0.550 to 0.657. Figure~\ref{fig:review_training} illustrates declining reconstruction losses in a separate first-window comparison; losses from the joint stages need not move together. A lower training loss therefore does not certify scenario fidelity.

These two paired windows support a narrow conclusion: improved reconstruction fit need not improve the cross-asset structure supplied to the optimizer. They do not show that 50 epochs are sufficient, rule out gains from additional training or hyperparameter changes, or characterize 200-epoch behavior throughout all 198 windows. The evaluated generator remains a specified fixed-budget implementation, not an estimate of the best attainable TimeGAN performance.

Appendix Figure~\ref{fig:supp_generation_variation} separates finite-generation variability with a fitted model held fixed from variability across fitted training seeds. At synthetic mass one-half, generating 1,512 rather than 756 rows leaves within-model L1 allocation variation near 22 percentage points, while between-model variation falls from 23.707 to 20.806 points. Sampling variation remains material under this restricted numerical budget; the comparison neither establishes a uniform count effect nor measures full-period investment performance.

\FloatBarrier

\section{Numerical Search and Reliability}\label{sec:numerics}

\subsection{Numerical Search and Supporting Linear Programs}\label{sec:numerical_search}
The ratio is nonconvex and nonsmooth. The common search uses sequential least squares programming (SLSQP) to minimize its negative, with analytic tail subgradients, at most 2,000 iterations and tolerance $10^{-10}$. Its 22 starts comprise the previous target, equal weights, all 12 corners and eight Dirichlet draws; numerical randomization is independent of the training seed. Initial points and finite returned candidates satisfying recorded simplex tolerances are compared after normalization. The previous-target initialization supplies an explicit fallback.

Independent supporting-function searches add candidates without changing the criterion. One searches the floor band through CVaR linear constraints; another uses fractional-programming supports on the negative-lower-tail branch~\citep{charnes1962programming}. Every proposal is rescored using Equation~\eqref{eq:weighted_rr} and retained only if its value exceeds the original search. A successful linear programming (LP) subproblem is not a global optimality certificate, and support candidates are not relabeled as successful SLSQP returns. The subproblems below state the two additional search regions explicitly.

Let $a_U$ and $a_L$ be asset-return vectors averaged with the frozen upper- and lower-tail coefficients from Section~\ref{sec:rrp}; their inner products with portfolio weights, less $r_f$, define the corresponding affine tail supports. The lower support is taken at a floor-feasible point. With a loss threshold $z$ and nonnegative excess-loss variables $u_j$, the floor-band subproblem takes the form
\begin{align*}
\max_{\boldsymbol w,z,\boldsymbol u}\quad &a_U^\top\boldsymbol w-r_f, \\
\text{subject to}\quad &u_j\geq-(\boldsymbol w^\top\boldsymbol r_j-r_f)-z,\quad u_j\geq0, \\
&z+\alpha^{-1}\sum_jp_ju_j\leq\varepsilon, \\
&a_L^\top\boldsymbol w-r_f\leq\varepsilon,\qquad\boldsymbol w\in\mathcal W.
\end{align*}
The CVaR constraint ensures that the true lower-tail mean is at least the negative floor. Since the true lower mean is the minimum over tail supports, the frozen-support cap ensures it is no greater than the positive floor. This is an inner search of the absolute-lower-tail floor band, not an economic exclusion of positive lower tails from the overall allocation rule. A maximum-lower-tail LP and zero-crossing calculation construct an anchor when the band is reachable. The fixed budget uses the original candidate, anchor and two upper-tail-ranked corners, with at most six support iterations. Small LP feasibility errors are recorded; any proposal is compared by its exact original criterion, without asserting a floor-band certificate for an infeasible LP approximation.

For the negative-lower-tail branch, a Charnes--Cooper representation~\citep{charnes1962programming} uses scaled weights and a nonnegative scale parameter:
\begin{align*}
&\boldsymbol y=\tau\boldsymbol w,\qquad \mathbf1^\top\boldsymbol y=\tau, \\
\max\quad&a_U^\top\boldsymbol y-r_f\tau, \\
\text{subject to}\quad &u_j\geq-\boldsymbol r_j^\top\boldsymbol y+r_f\tau-z,\qquad u_j\geq0, \\
&z+\alpha^{-1}\sum_jp_ju_j\leq1, \\
&a_L^\top\boldsymbol y+(\varepsilon-r_f)\tau\leq0,\qquad\boldsymbol y\geq0,\quad\tau\geq0.
\end{align*}
Supports are updated from returned weights, initialized from the original selected candidate and equal weights, with at most six iterations. The branch is an additional search region; the overall rule retains the best original or new candidate over its full original domain. The main procedure uses the original floor. Section~\ref{sec:floor_sensitivity} instead changes the denominator floor while retaining the original feasible weight set; it is a criterion sensitivity, not an additional search for the baseline objective.

\subsection{Denominator Behavior and Economic Interpretation}\label{sec:denominator}

\begin{figure}[!htbp]
\centering
\includegraphics[width=.84\textwidth,height=.58\textheight,keepaspectratio]{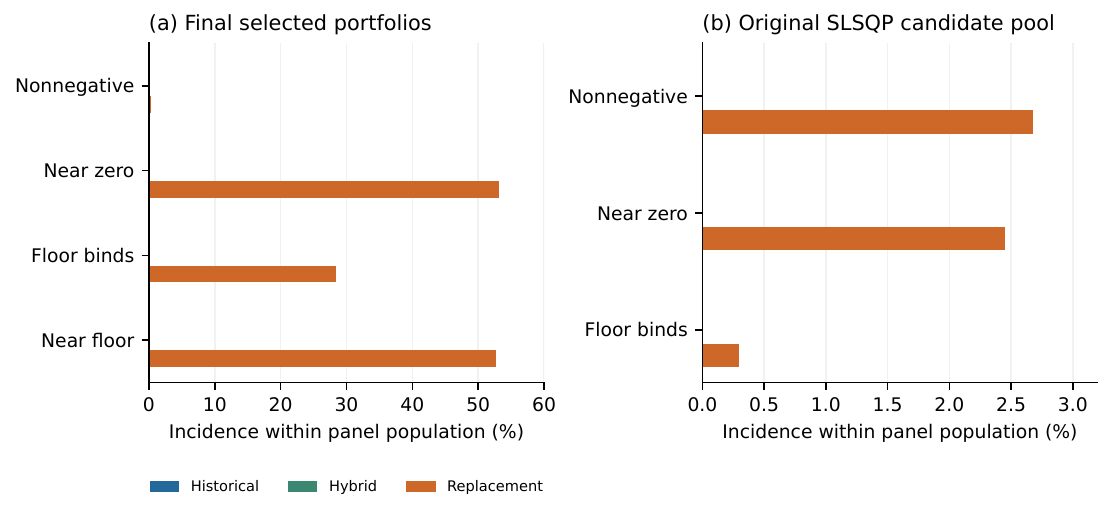}
\caption{Denominator diagnostics for distinct populations. Panel (a) uses final selected decisions: 198 historical, 3,960 hybrid and 1,980 replacement. Panel (b) uses 8,712, 174,240 and 66,257 original SLSQP candidate points, respectively, including repeated initial and feasible returned points and excluding extra support proposals. Nonnegative means lower-tail mean at least zero; near zero uses absolute value at most $10^{-6}$; floor binding uses $10^{-8}$; near floor allows factor 1.0001. Categories overlap. Panel scales differ; percentages are not shares of one common pool.}\label{fig:denominator_populations}
\end{figure}

Figure~\ref{fig:denominator_populations}(a) shows that 53.23\% of selected replacement portfolios have an absolute lower-tail mean at most $10^{-6}$ and 28.38\% strictly bind the $10^{-8}$ floor; both incidences are zero for historical and hybrid selections. Figure~\ref{fig:denominator_populations}(b) reports a different population: only 2.45\% of original replacement SLSQP candidates are near zero and 0.30\% strictly bind the floor. The difference makes it essential to distinguish the searched candidate pool from final selections, which can also come from supporting-function searches; the two percentages are not estimates over a common population.

More broadly, 1,045 of 1,980 replacement decisions have an absolute lower-tail mean no greater than 1.0001 times the floor; no historical or hybrid decision satisfies that near-floor criterion. Independent candidates can raise in-window ratios above $10^5$ without establishing exceptional subsequent returns. The lower tail of a generated portfolio can be close to zero even though realized market losses remain material. This combination concerns the interaction of the estimated distribution and modified objective, rather than one identifiable neural-network programming error.

An economic-scale floor is a different criterion, not a numerical repair. Section~\ref{sec:floor_sensitivity} tests that criterion change on fixed current-model inputs with the same fractional-tail and support-search procedure. The full-period comparison identifies sensitivity of the implemented policies, without justifying selection of a floor from its realized payoff. Its near-own-floor indicator uses a tolerance and differs from the strict-floor and near-zero counts reported here. The criterion, scenario distribution and numerical stopping evidence must be stated together.

\FloatBarrier

\subsection{Solver Diagnostics and Candidate Selection}\label{sec:numerical_reliability}
\begin{figure}[!htbp]
\centering
\includegraphics[width=.86\textwidth,height=.58\textheight,keepaspectratio]{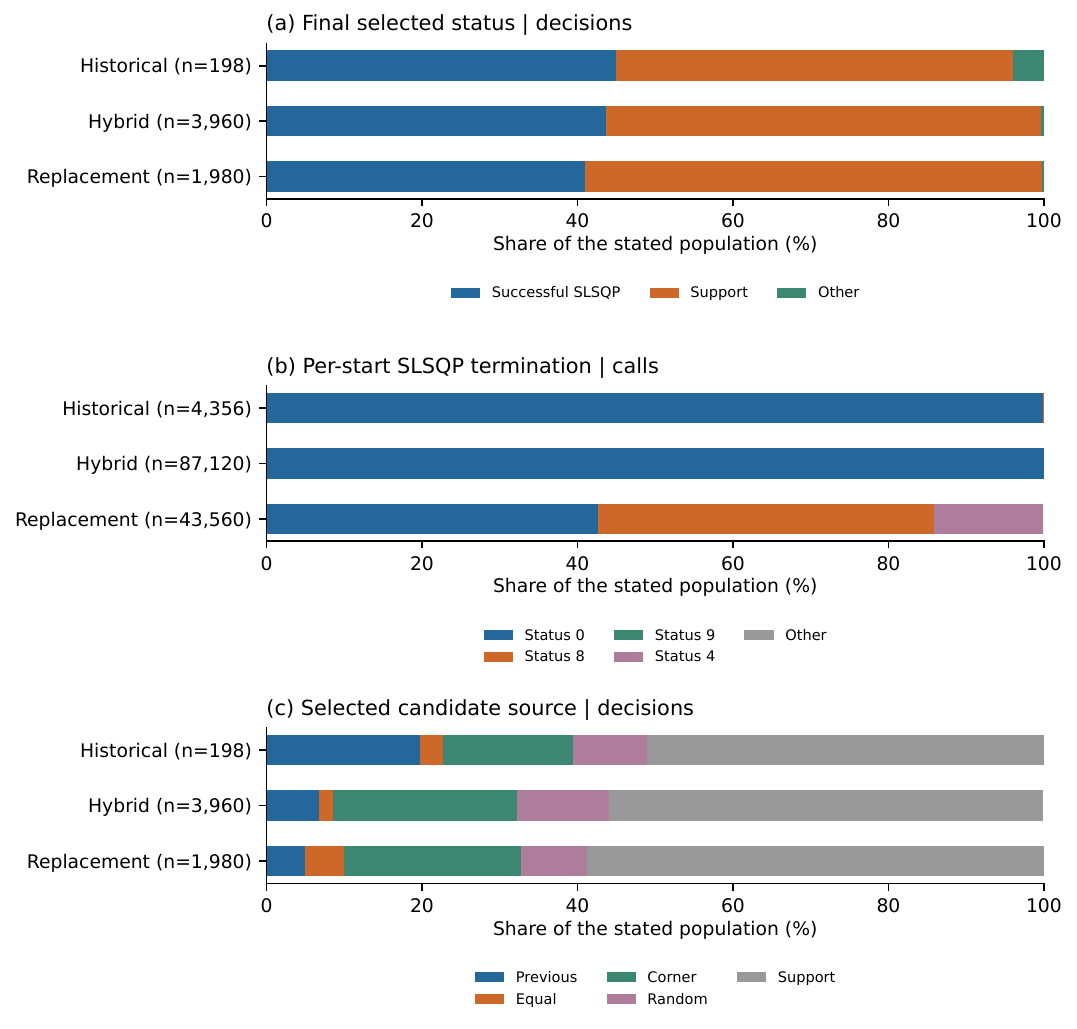}
\caption{Numerical outcomes by population. Panel (a) classifies the final selection; (b) classifies individual original SLSQP calls; (c) identifies selected-candidate sources. Each row gives its own denominator. SLSQP status 0 denotes successful termination, 8 positive directional derivative, 9 iteration limit and 4 incompatible inequalities. Percentages may not sum to 100 after rounding. Support selections have no global-optimality certificate; per-call success does not classify final support candidates.}\label{fig:solver_populations}
\end{figure}

The completed experiment contains 135,036 original SLSQP starts. Figure~\ref{fig:solver_populations}(a) separates selected numerical outcomes: 2,629 decisions select successful SLSQP returns, 3,480 select support candidates, and 29 retain other labels. Figure~\ref{fig:solver_populations}(b) instead counts individual SLSQP calls: successful termination is about 99.9--100.0\% for historical and hybrid calls but only 42.6\% for replacement calls; replacement also reports 43.3\% positive-directional-derivative and 14.0\% incompatible-inequality flags. Such flags are not proof that the feasible economic weight set is empty. Figure~\ref{fig:solver_populations}(c) shows that supporting-function candidates supply 51.0\%, 55.9\% and 58.8\% of historical, hybrid and replacement selections, respectively. Their substantial role explains why per-call SLSQP success cannot stand in for final-candidate reliability or a global-optimality certificate. The 29 unsuccessful selections include 24 retained previous targets with successful alternatives differing by at most about $3.4\times10^{-8}$ in relative objective, and five replacement selections without a successful returned start. These flags remain numerical limitations.

\FloatBarrier

\section{Investment Outcomes and Conditional Evidence}\label{sec:investment_results}

\subsection{Net Performance and Seed Dispersion}\label{sec:investment}
The matched comparison contains a common historical reference, 20 hybrid policies and ten replacement policies, each with 198 decisions and 3,955 daily returns. All use the same execution dates, constraints and cost convention. Performance measures follow Section~\ref{sec:performance_measures}.

\begin{figure}[!htbp]
\centering
\includegraphics[width=.84\textwidth,height=.58\textheight,keepaspectratio]{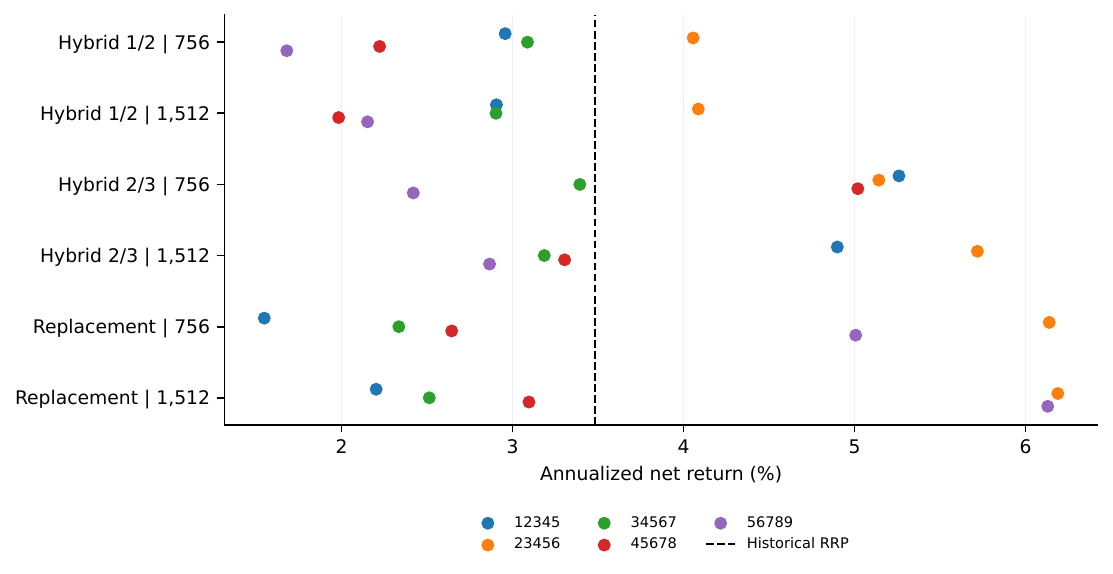}
\caption{Annualized net returns of the five fixed training-seed policies in each mass--count configuration, with the common historical reference. Every colored point is one policy statistic, not an ensemble return. Small vertical offsets separate seeds. The spread is descriptive variation across fitted policies on the same history, not a confidence interval or five independent market replications.}\label{fig:seed_dispersion}
\end{figure}

Table~\ref{tab:new_performance} reports the completed comparison. Generated-policy cells are means of the five seed-specific performance statistics; they are not the performance of a portfolio formed by averaging those policies. Figure~\ref{fig:seed_dispersion} displays every fitted seed; Section~\ref{sec:allocations} examines portfolio concentration.

\begin{table}[!htbp]
\centering\small\setlength{\tabcolsep}{3.5pt}
\caption{Net performance under the completed experiment}\label{tab:new_performance}
\begin{tabularx}{\textwidth}{@{}Xrrrrrrrr@{}}
\toprule
Procedure & $M$ & Ret. & Vol. & SR & RR & MDD & CVaR & TO \\
\midrule
Historical RRP & -- & 3.48 & 9.42 & 0.26 & 0.96 & 39.59 & 1.45 & 234.29 \\
Equal-weight & -- & 6.41 & 10.62 & 0.50 & 0.92 & 23.42 & 1.58 & 17.32 \\
Buy-and-hold & -- & 8.41 & 11.77 & 0.62 & 0.91 & 25.66 & 1.77 & 0.00 \\
\midrule
Hybrid $\eta=1/2$ & 756 & 2.80 & 8.98 & 0.19 & 0.94 & 41.50 & 1.41 & 420.65 \\
Hybrid $\eta=1/2$ & 1512 & 2.81 & 8.98 & 0.20 & 0.94 & 41.49 & 1.40 & 403.75 \\
\midrule
Hybrid $\eta=2/3$ & 756 & 4.25 & 9.45 & 0.34 & 0.97 & 40.19 & 1.46 & 551.07 \\
Hybrid $\eta=2/3$ & 1512 & 3.99 & 9.50 & 0.31 & 0.96 & 39.87 & 1.48 & 539.31 \\
\midrule
Replacement & 756 & 3.53 & 13.17 & 0.22 & 0.91 & 39.43 & 2.10 & 847.97 \\
Replacement & 1512 & 4.03 & 13.09 & 0.26 & 0.93 & 36.80 & 2.07 & 844.62 \\
\midrule
\multicolumn{9}{@{}p{\textwidth}@{}}{\footnotesize\textit{Note:} Return, volatility and one-way turnover are annualized percentages; MDD and daily 5\% loss-CVaR are percentages. SR is annualized; RR is the unitless fractional realized ratio. Cost is 10 bp per total value traded. Each generated cell averages five fixed-seed statistics; passive rules have no fitted-model seed.} \\
\bottomrule
\end{tabularx}
\end{table}

In Table~\ref{tab:new_performance}, historical RRP earns 3.48\% annualized net return, versus about 2.80\% for either equal-mass hybrid count. The higher-mass hybrid means are 4.25\% with 756 draws and 3.99\% with 1,512 draws. Replacement means are 3.53\% and 4.03\%, respectively. Figure~\ref{fig:seed_dispersion} shows that higher-mass hybrid returns span 2.42--5.26\% at 756 rows and 2.87--5.72\% at 1,512 rows, so the favorable configuration averages are not shared uniformly across fits. Figure~\ref{fig:seed_dispersion} likewise shows replacement returns spanning 1.55--6.14\% at 756 rows and 2.20--6.19\% at 1,512 rows. Only two of five replacement seeds exceed the historical return at either count. Higher-mass hybrids exceed it in three seeds at 756 rows and two at 1,512 rows. Increasing count raises replacement return in all five seeds, while hybrid count effects are mixed. The nested prefixes pair these comparisons within each fitted model; persistent fidelity errors and conditional uncertainty prevent a general claim of benefit from more simulated rows.

The passive references have greater net growth: 6.41\% for equal-weight rebalancing and 8.41\% for buy-and-hold; none of the 30 generated policies exceeds either in this sample. Their greater net growth does not by itself establish superior tail-ratio performance. The historical realized ratio is about 0.956, compared with 0.909 for buy-and-hold. Reward-risk performance and geometric growth should therefore remain separate statements.

\begin{figure}[!htbp]\centering
\includegraphics[width=.86\textwidth,height=.58\textheight,keepaspectratio]{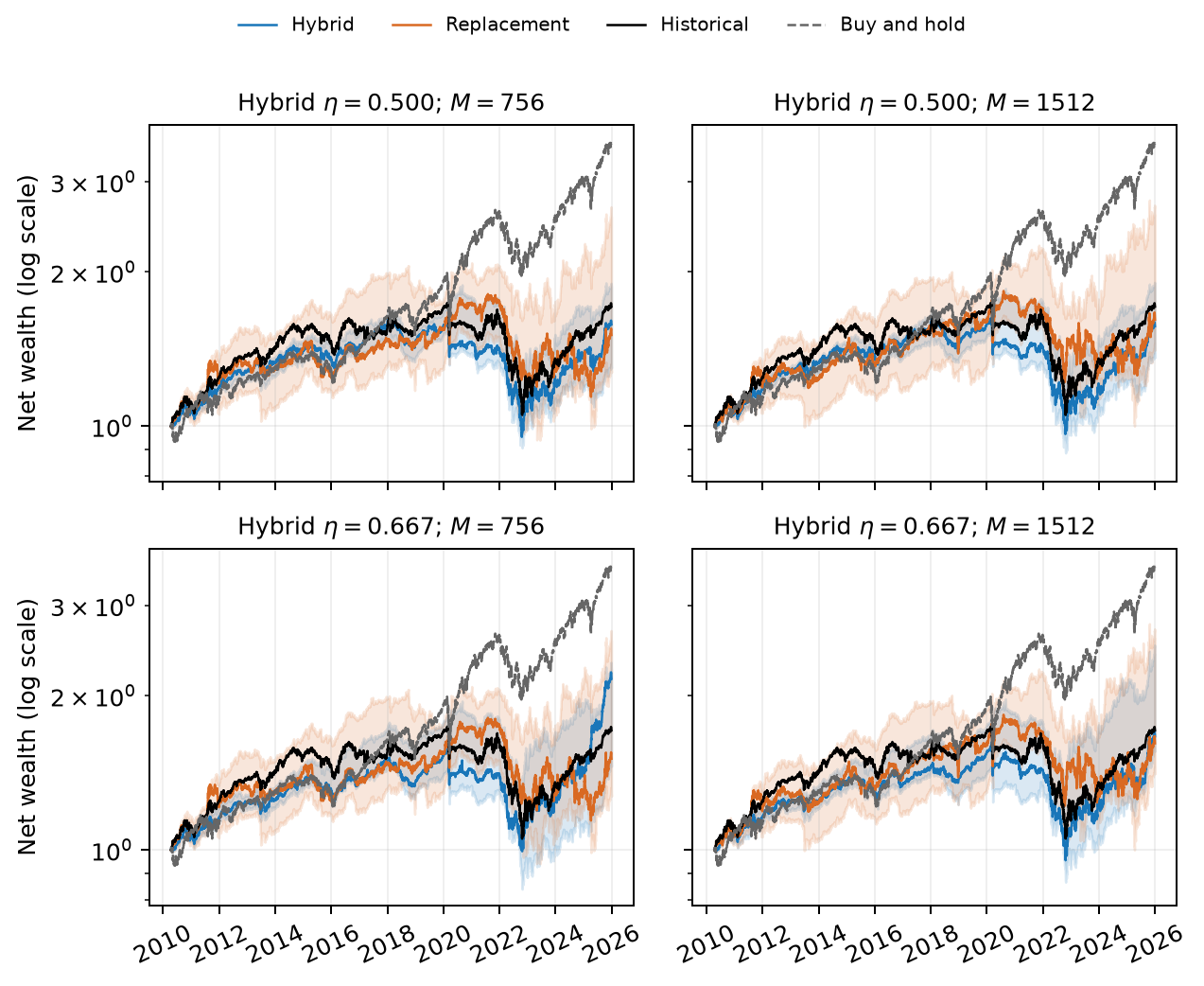}
\caption{Net wealth of the completed policies. Colored lines are pointwise medians and shaded regions are the minimum--maximum range across five fixed seeds. These descriptive curves are not implementable averaged portfolios. Historical RRP and passive buy-and-hold are common references; all series use the same earned-return dates and applicable trading costs.}\label{fig:new_wealth}
\end{figure}

Replacement incurs substantially more trading: its mean annual one-way turnover is roughly 840--850\%, versus 234\% for historical RRP. Mean one-way turnover per scheduled decision is approximately 67\% for replacement and 19\% for the historical reference, including the uncharged initial formation. Hybrids also trade more than the historical reference. Neither a larger in-window ratio nor a favorable gross-return comparison establishes net economic dominance. Figure~\ref{fig:new_wealth} shows the common buy-and-hold path ending above all generated-policy bands, while replacement bands are substantially wider than the equal-mass hybrid bands. The higher-mass hybrid paths separate more strongly late in the sample. These pointwise median lines and minimum--maximum bands describe variation across fixed fitted policies; they are neither confidence bands nor the wealth of a combined investable portfolio.

\FloatBarrier

\subsection{Allocations, Concentration, and Trading}\label{sec:allocations}
\begin{figure}[!htbp]
\centering
\includegraphics[width=.90\textwidth,height=.58\textheight,keepaspectratio]{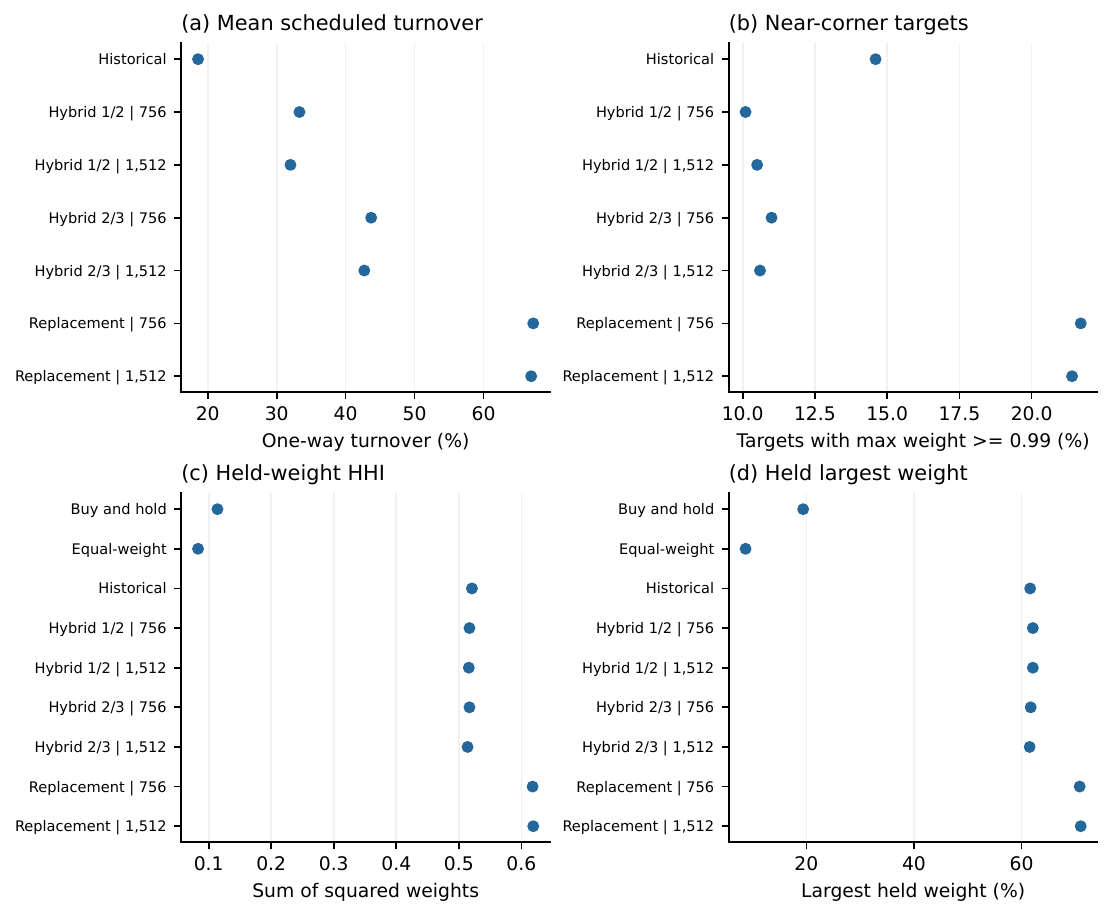}
\caption{Trading and concentration under the main experiment. Turnover averages all 198 scheduled decisions, including zero at initial formation. Near-corner frequency uses target maximum weight at least 0.99. HHI and largest weights use daily held weights after trading and price drift. Generated configurations average five seed-specific statistics. HHI is unitless; the other panels are percentages. Passive references appear only in held-weight panels.}\label{fig:trading_concentration}
\end{figure}

\begin{figure}[!htbp]\centering
\includegraphics[width=.86\textwidth,height=.58\textheight,keepaspectratio]{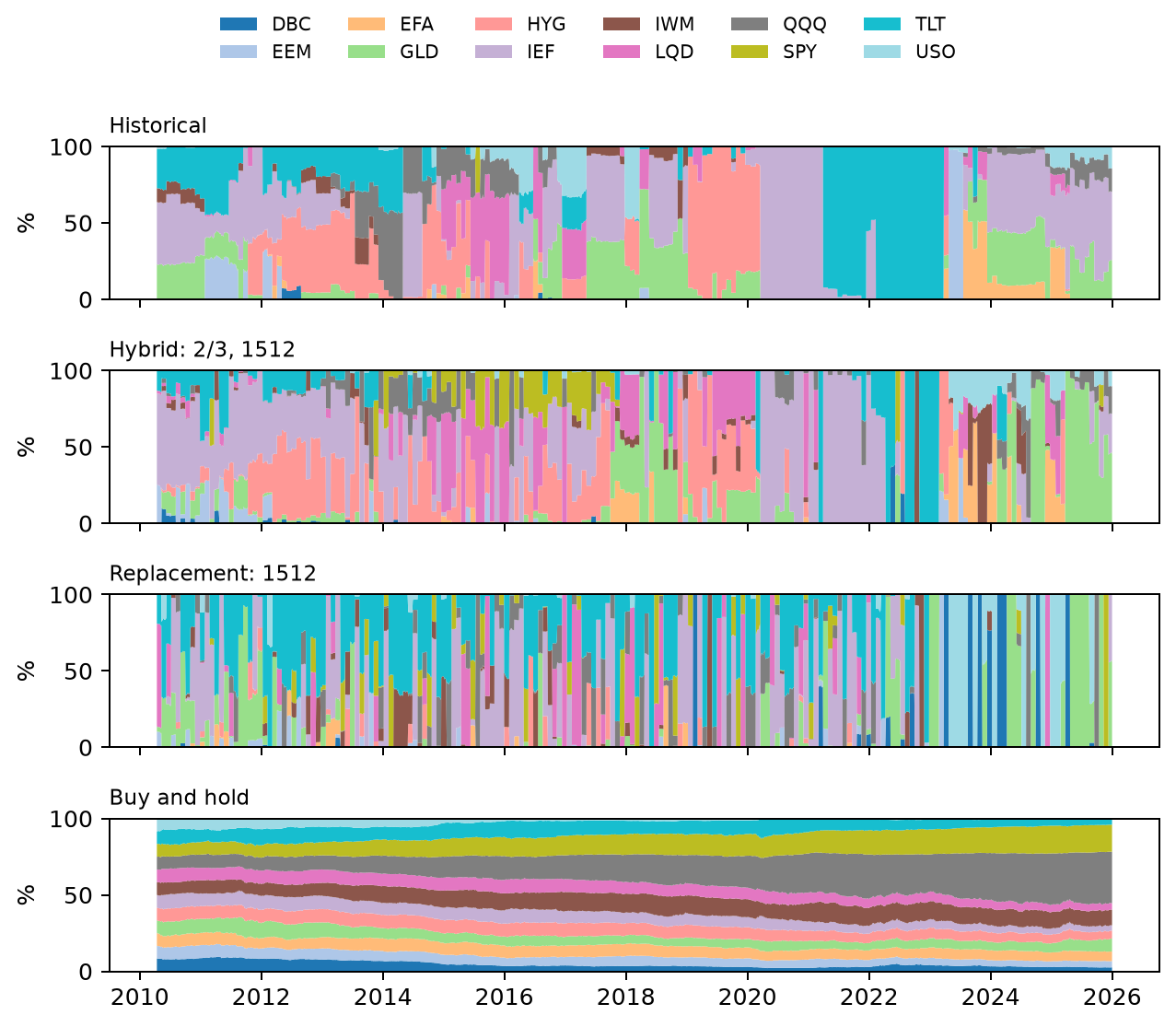}
\caption{Daily held asset weights for the common historical reference, primary-seed higher-mass hybrid and replacement policies at 1,512 rows, and passive buy-and-hold. Intervening movements include price drift.}\label{fig:s_weights}
\end{figure}

Figure~\ref{fig:trading_concentration} shows mean one-way turnover of 66.9--67.2\% per scheduled replacement decision, compared with 18.6\% for historical allocation; replacement also has about 21\% near-corner targets, versus 14.6\% historically. Figure~\ref{fig:trading_concentration} measures daily held-weight concentration with the Herfindahl--Hirschman index (HHI), including intervening drift. Replacement HHI is 0.618--0.619, above 0.521 historically and roughly 0.514--0.517 for hybrids, and its mean largest weight is about 71\%. EWP remains near the equal-weight HHI of $1/12$. Figure~\ref{fig:s_weights} provides the corresponding time-path view: the illustrative replacement policy switches exposures more frequently, whereas historical allocation holds some dominant bond exposures for extended periods and buy-and-hold weights change smoothly through price drift. This separates active trading from passive concentration changes.

\begin{figure}[!htbp]
\centering
\includegraphics[width=.78\textwidth,height=.58\textheight,keepaspectratio]{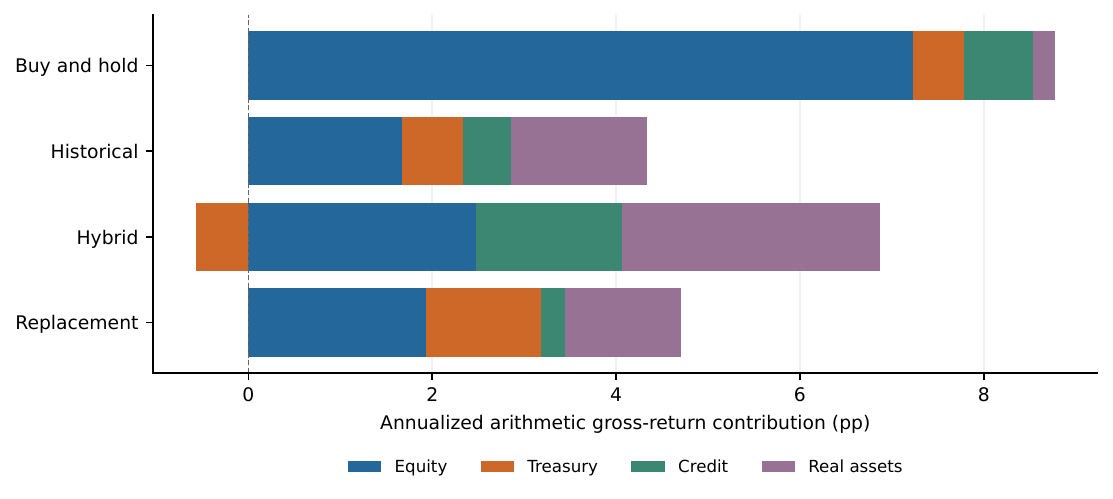}
\caption{Annualized arithmetic gross-return contributions by asset group, before portfolio costs. Positive and negative components stack separately. Groups are the five equity ETFs, IEF/TLT, HYG/LQD and DBC/GLD/USO. Generated illustrations use seed 12345, 1,512 rows and hybrid mass $2/3$. These contributions sum to arithmetic gross return, not geometric net growth.}\label{fig:asset_group_contributions}
\end{figure}

Appendix Figure~\ref{fig:supp_allocation} shows that buy-and-hold starts equally weighted but ends with approximately 33.7\% in QQQ, 17.7\% in SPY, and 3.6\% in IEF. Its growth ranking reflects this drift as well as the lack of subsequent trading. The asset-group contributions in Figure~\ref{fig:asset_group_contributions} are annualized arithmetic gross-return contributions, not a decomposition of geometric net wealth. Figure~\ref{fig:asset_group_contributions} attributes 7.23 annualized arithmetic gross-return percentage points to buy-and-hold equities, compared with 1.67 for the historical policy. The illustrative hybrid has a negative Treasury contribution of -0.57 points but positive credit and real-asset contributions of 1.58 and 2.81 points. This accounting separates exposures contributing to returns without identifying a causal explanation or reconstructing geometric net growth.

\FloatBarrier

\subsection{Conditional Uncertainty and Block-Length Sensitivity}\label{sec:inference}

Synchronized moving-block bootstrap samples calendar blocks of the realized net-return and lagged-rate series~\citep{kunsch1989jackknife}. We use 2,000 replicates and block lengths 10, 20 and 40. Each replicate uses identical dates for every policy, preserving paired comparisons. The reference length is 20, matching the decision spacing; the other lengths assess sensitivity, not stationarity across market regimes. Pooling short calendar blocks provides an approximation to short-range dependence in these fixed return paths. It does not explicitly model regime transitions or assert that the entire 2010--2025 distribution is stable; block-length and fixed-calendar episode results make this limitation visible.

For each seed-specific policy, intervals concern the difference from the historical reference. For a configuration, we average the five fixed-seed performance statistics within each synchronized resample. This does not treat five seeds as independent market histories or construct an ensemble portfolio. Percentile 95\% intervals are pointwise.

Appendix Figure~\ref{fig:appendix_block_sensitivity} provides the 10- and 40-day interval sensitivity.

\begin{figure}[!htbp]
\centering
\includegraphics[width=.92\textwidth,height=.58\textheight,keepaspectratio]{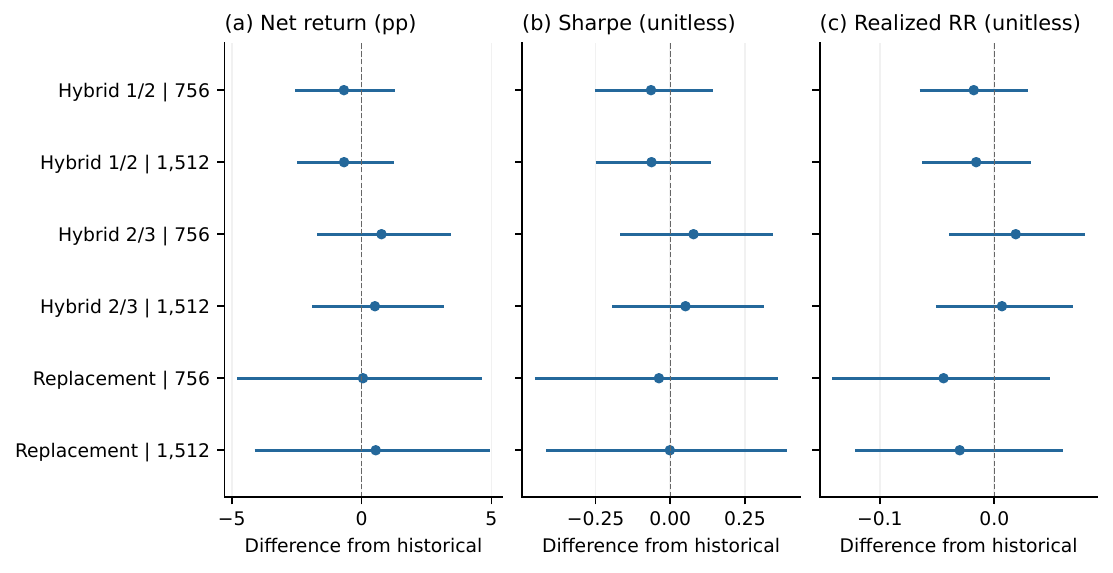}
\caption{Conditional configuration-average differences from historical RRP in annualized net return, Sharpe ratio and realized fractional tail ratio. Points average five fixed-seed statistics; bars are pointwise 95\% percentile intervals from 2,000 synchronized overlapping 20-day-block resamples. Each metric has its own units and scale. All panels condition on the fitted policies and exclude training, reoptimization and specification-selection uncertainty. The 10- and 40-day sensitivity intervals appear in Appendix Figure~\ref{fig:appendix_block_sensitivity}.}\label{fig:new_intervals}
\end{figure}

At the reference block length, Figure~\ref{fig:new_intervals} shows that intervals include zero for all six configuration-average return contrasts; the Sharpe and realized-ratio intervals also include zero. They also include zero for all 30 individual generated-versus-historical comparisons on each of those metrics. The evidence does not establish a stable advantage. It also does not establish equivalence, absence of economically relevant effects, or an upper bound on attainable performance of other generators.

The inference is conditional on completed policy paths. It does not refit the generator, repeat optimization, reconstruct holdings/costs inside resamples, or integrate uncertainty from selecting the asset universe and research specification. Resampling fixed returns is therefore narrower than an unconditional test of a learning-and-allocation procedure. Appendix Figure~\ref{fig:appendix_block_sensitivity} shows that all configuration-average return, Sharpe and realized-RR intervals include zero at the 10- and 40-day block lengths. Figure~\ref{fig:new_intervals} presents the 20-day reference results on separate metric scales. Interval endpoints vary with block length, especially for replacement, but none of these settings establishes a configuration-average advantage on the three reported metrics.

Figure~\ref{fig:new_intervals} shows all six configuration-average return intervals crossing the zero line at the reference 20-day block length. The wider replacement intervals visualize greater conditional uncertainty than the hybrid intervals, even where the point estimates are positive. The historical-minus-buy-and-hold comparison in Appendix Figure~\ref{fig:supp_passive} further illustrates block sensitivity: its 20-day-block upper endpoint is approximately -0.000376 annualized percentage points, while the 10-day-block interval includes zero. Appendix Figure~\ref{fig:supp_passive} also shows that historical-minus-EWP return intervals and both passive-reference RR contrasts include zero at all three block lengths. The near-zero EWBH endpoint and its block-length sensitivity qualify the inferential interpretation of the passive growth advantage.

The electronic inference tables include all 30 individual comparisons against historical RRP, 15 paired count contrasts, and ten probability-mass contrasts. Identical policies have exactly zero paired differences in every resample. Reference-length intervals against historical RRP include zero for all individual policies and all three reported metrics. Centered-bootstrap two-sided tail frequencies are exploratory; Holm adjustment is applied separately within a contrast family, metric and block length. The different block lengths are sensitivity checks, not three independent confirmations.

Fixed-policy inference samples 2,000 synchronized overlapping calendar-block paths without circular wrap. It does not reconstruct a feasible trading strategy on each resampled calendar, refit networks, repeat optimization or rebuild transaction costs. These limitations apply even when a pointwise interval excludes zero. The analysis does not infer equivalence from intervals containing zero.

\subsection{Cost Repricing and Fixed-Calendar Episodes}\label{sec:costs_episodes}
\begin{figure}[!htbp]
\centering
\includegraphics[width=.82\textwidth,height=.58\textheight,keepaspectratio]{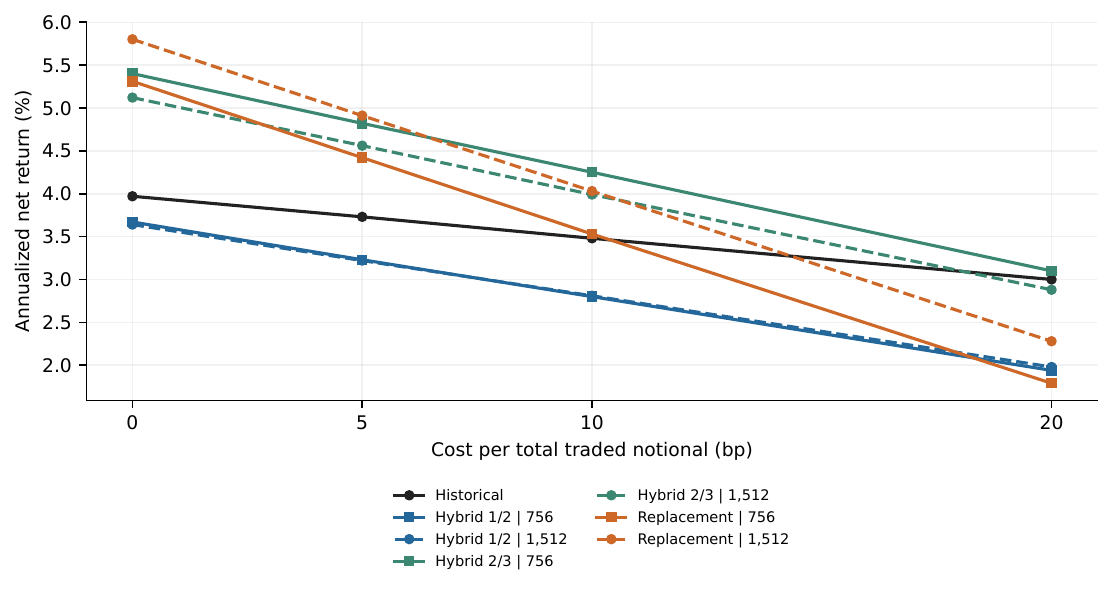}
\caption{Fixed-allocation cost repricing at 0, 5, 10 and 20 bp per total traded notional. Gross return paths and targets are unchanged; generated values average the five fixed-seed statistics. Lines connect the four evaluated fee levels and do not represent cost-aware reoptimization.}\label{fig:cost_repricing}
\end{figure}

Trading-cost sensitivity is repriced from saved gross returns and realized turnover at 0, 5, 10 and 20 bp, leaving targets fixed. It is an accounting sensitivity, not a cost-aware reoptimization. Figure~\ref{fig:cost_repricing} shows historical annualized return falling from 3.97\% at zero cost to 3.00\% at 20 bp, while replacement falls from 5.31\% to 1.79\% at 756 rows and from 5.80\% to 2.28\% at 1,512 rows. Thus, the larger replacement turnover translates into a larger fee burden on these fixed paths. Appendix Figure~\ref{fig:supp_episodes} shows that the relative ordering also varies across calendar slices: replacement earns about 5.2--5.3\% in 2020--2021 while historical and hybrid configurations have negative returns; in 2022--2025 the higher-mass hybrids earn about 5.7--7.0\%, compared with 1.24\% historically. These shifts help explain why a whole-period average does not describe every market episode. Subperiods are descriptive slices of the same realized path, not independent replications or crisis-timing rules selected for favorable outcomes.

\FloatBarrier

\section{Robustness, Benchmarks, and Limitations}\label{sec:robustness}

The full-period comparisons use all 198 decision dates rather than the ten model windows previously selected by generation error. A separate 62-policy calculation reuses the original fitted TimeGAN draws and frozen market data: two historical floor variants, 30 TimeGAN mass--floor policies and 30 IID or moving-block policies. All use 756 additional rows when applicable, exact fractional 5\% tails, the long-only simplex and identical numerical budgets. Each policy recursively uses its own previous target and drifted pre-trade holdings, with delayed execution and the original trading-cost rule. These are alternative policies on the same market history, not independent market replications or additional model fits.

Every baseline and sensitivity arm is recomputed in one local runtime. The historical reference earns 3.53\% annualized net return here, versus 3.48\% in the accepted main experiment; small numerical differences can propagate through recursive allocations. Accordingly, comparisons in this section use the new common baseline, while Section~\ref{sec:investment_results} retains the original 31-policy count--mass experiment. The full design independently verifies 12,276 decision objectives, holdings and return accounting. Earlier ten-window switch diagnostics remain in the reproducibility records but do not determine the full-period source comparison.

\subsection{Sensitivity to the Modified Tail-Ratio Criterion}\label{sec:floor_sensitivity}

The baseline criterion retains the $10^{-8}$ denominator floor in Equation~\eqref{eq:weighted_rr}. The alternative is fixed from the current historical window before subsequent returns are observed:
\begin{equation*}
\varepsilon_t^{\mathrm{scale}}=\max\{10^{-6},\ 0.1\,\mathrm{ETL}_{0.05}^{\mathrm{hist}}(\boldsymbol w^{\mathrm{EW}};t)\}.
\end{equation*}
Here $\boldsymbol w^{\mathrm{EW}}$ is the equal-weight portfolio and ETL uses historical excess returns with fractional tail weights. The same rule applies to historical, hybrid and replacement inputs. The multiplier is not selected by subsequent investment returns. This changes the implemented preference criterion; it neither repairs the generated distribution nor supplies a neutral numerical correction.

\begin{figure}[!htbp]
\centering
\includegraphics[width=.84\textwidth,height=.58\textheight,keepaspectratio]{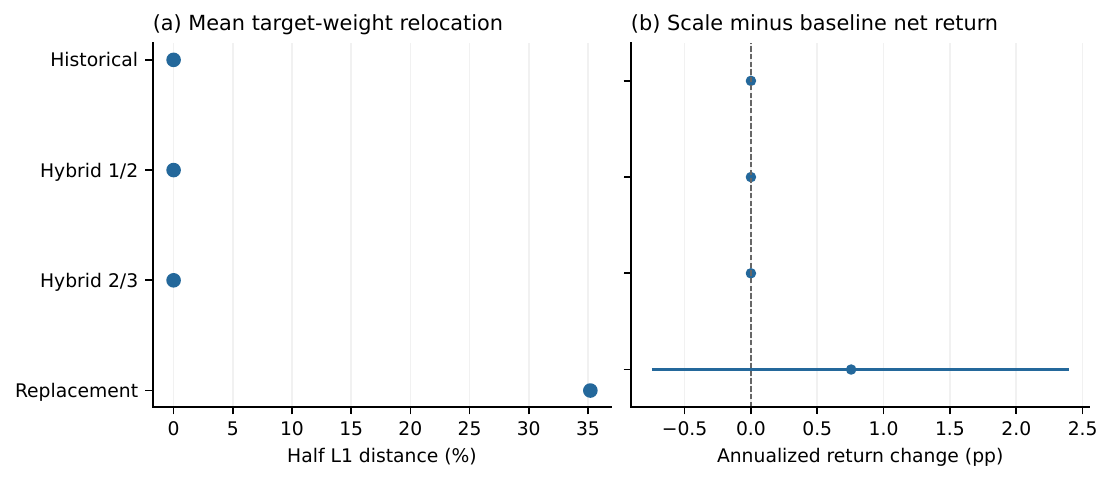}
\caption{Full-period sensitivity to the historical-scale denominator at 756 generated rows. Panel (a) reports the mean half-L1 distance between contemporaneous targets; (b) reports scale-minus-baseline annualized net-return contrasts with pointwise conditional 95\% intervals from 2,000 synchronized overlapping 20-day-block resamples. Historical comparisons cover 198 decisions; generated comparisons cover five seeds and 990 decisions. Policies follow their own recursive holdings and costs. Different floors define different objective scales; no objective-value improvement is inferred. Incidence and objective summaries are discussed in the text; underlying records are retained for reproducibility.}\label{fig:floor_sensitivity}
\end{figure}

Figure~\ref{fig:floor_sensitivity} pairs baseline and scale-floor policies at every decision date. A near-own-floor decision has an absolute lower-tail mean no greater than 1.0001 times that policy's own floor. This differs from both the strict-floor and $10^{-6}$ near-zero indicators in Section~\ref{sec:denominator}. Half the L1 difference between contemporaneous targets measures weight relocation; it includes any propagation through previous targets. Objective values under the two floors are on different criterion scales and are not ranked as investment improvements.

Figure~\ref{fig:floor_sensitivity} shows historical and hybrid allocations unchanged to numerical precision and replacement relocating 35.21\% of target weight on average. Its median in-window objective falls from $5.74\times10^4$ to 1.74. Near-own-floor counts change from 525 to 762 out of 990 decisions; this increase uses different thresholds and selected weights, not a common definition of proximity to zero. Mean annualized net return rises from 3.51\% to 4.27\%, but the paired change is +0.76 [-0.75, 2.39] percentage points. The interval includes zero at each of the 10-, 20- and 40-day block lengths.

The scale floor does not uniformly improve realized risk or reduce the observed seed dispersion. Replacement's configuration-average daily loss-CVaR increases from 2.10\% to 2.29\%, while seed returns span 0.88--8.89\% rather than 1.67--6.01\%. Thus, a smaller optimized ratio is not itself evidence of better downside protection or a superior floor choice. These full-period policy sensitivities extend the earlier local switch diagnostics, but do not identify a causal decomposition of the generator's contribution to investment performance.

\FloatBarrier

\subsection{Comparison with Simple Scenario Resampling}\label{sec:resampling}

IID and moving-block controls compare learned scenarios with resampling observed returns. Each sampled observation is a complete 12-asset daily return vector, preserving the contemporaneous cross-asset values in that row. IID sampling draws 756 historical indices with replacement. Circular moving-block sampling draws uniformly selected starts, takes 20 consecutive rows per block with wrap-around inside the historical window, and truncates to 756 rows. Both use fixed seed rules and the same probability masses as TimeGAN. All source comparisons retain the baseline denominator, common decision dates and identical search budgets.

\begin{figure}[!htbp]
\centering
\includegraphics[width=.96\textwidth,height=.58\textheight,keepaspectratio]{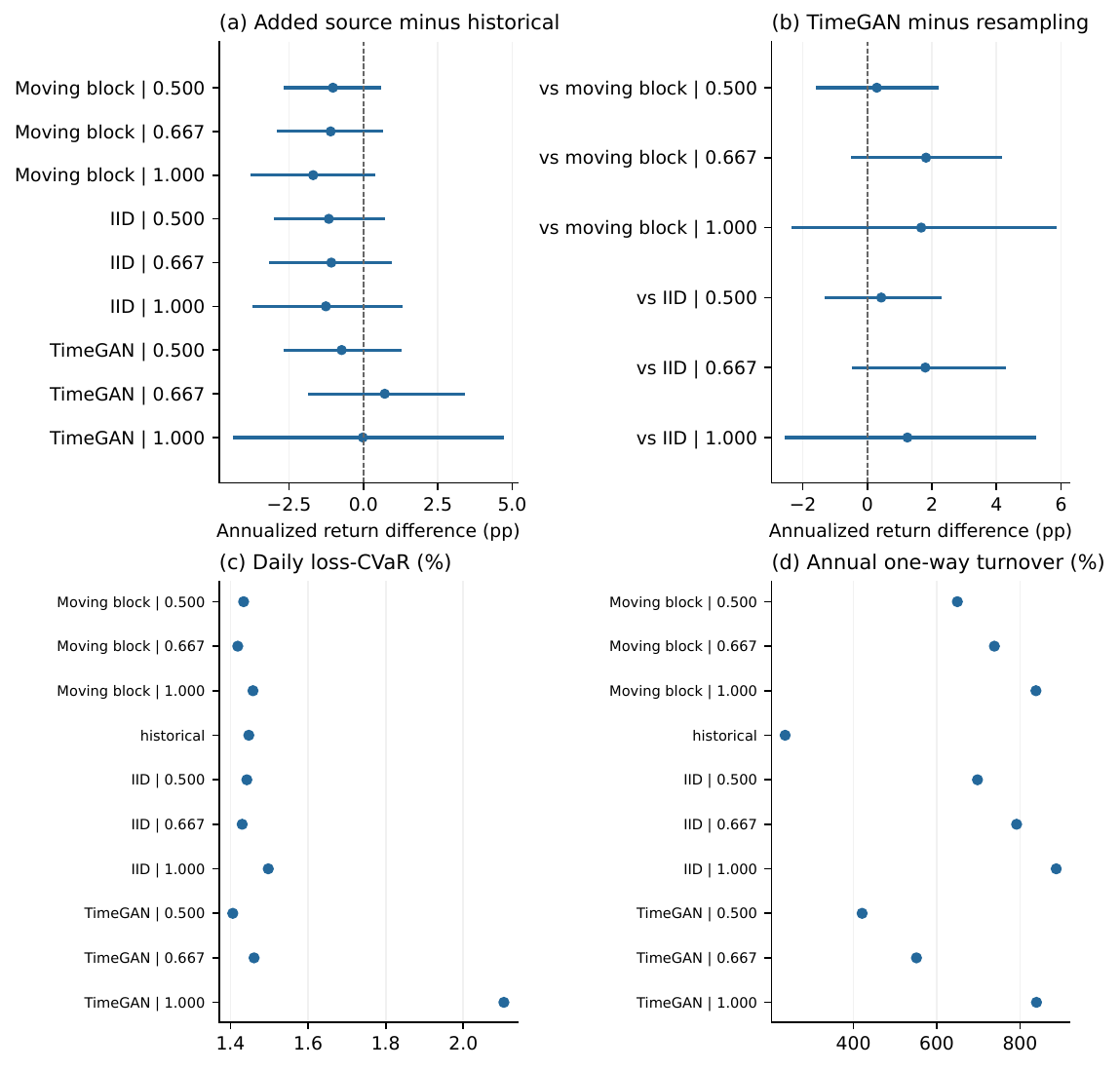}
\caption{Full-period source comparisons at 756 added rows. Panel (a) compares each source with the common local-runtime historical replay; (b) directly compares TimeGAN with IID and moving-block resampling. Labels give added probability mass. Points and bars are annualized net-return contrasts and pointwise conditional 95\% intervals from 2,000 synchronized overlapping 20-day-block resamples. Panels (c)--(d) give five-seed-average daily net-return loss-CVaR and annual one-way turnover, with one shared historical reference. This replay is distinct from the main frozen count comparison. Intervals condition on saved policies and exclude refitting and reoptimization.}\label{fig:scenario_source_comparison}
\end{figure}

Figure~\ref{fig:scenario_source_comparison} summarizes recursively implemented policy contrasts, risk and trading. The reproducibility records retain the underlying performance statistics. Returns, realized ratios and losses follow Section~\ref{sec:performance_measures}; return contrasts use the historical replay in this section. The bootstrap resamples identical calendar blocks across all saved paths, with 2,000 replicates and block lengths 10, 20 and 40, as in Section~\ref{sec:inference}. Configuration contrasts average the five fixed-seed performance statistics rather than construct an investable ensemble. Intervals condition on fitted policies and omit refitting, reoptimization and specification selection.

Figure~\ref{fig:scenario_source_comparison} displays the corresponding contrasts. TimeGAN net-return means are 2.80\%, 4.25\% and 3.51\% at masses $1/2$, $2/3$ and 1, respectively. The IID means are 2.37\%, 2.45\% and 2.27\%; moving-block means are 2.51\%, 2.43\% and 1.84\%. All nine source-versus-historical return intervals include zero at the reference length. At mass $2/3$, the direct TimeGAN-minus-IID return contrast is +1.80 [-0.48, 4.28] percentage points, and TimeGAN-minus-moving-block is +1.82 [-0.51, 4.18]. All six direct TimeGAN--resampling comparisons have return, Sharpe and realized-RR intervals containing zero at all three block lengths. The mean-return ordering therefore does not establish persistent superiority; zero-containing intervals also do not establish equivalence.

The risk and trading results qualify that ordering. Full TimeGAN replacement has daily loss-CVaR of 2.10\%, versus 1.50\% for IID, 1.46\% for moving blocks and 1.45\% historically. TimeGAN hybrid turnover is lower than the corresponding resampling controls, but replacement turnover remains about 840\% annually, versus 236\% historically. The electronic results retain every paired contrast, metric and block length, including exploratory centered-bootstrap tail frequencies with Holm adjustment within contrast family, metric and block length; no favorable seed or metric determines the comparison.

For the pooled one-period objective, permuting rows leaves the empirical distribution and criterion unchanged. Moving blocks change the sampled observations and multiplicities; their retained order is not directly used by the optimizer. Neither resampling nor additional draws create newly observed tail events. The full-period controls consequently assess the implemented scenario distributions and sampling rules on this shared history, rather than prove that temporal generation itself adds economic information.

\FloatBarrier

\subsection{Discussion and Limitations}\label{sec:limits}\label{sec:design_sensitivity}

The findings distinguish statistical fidelity from suitability for tail-sensitive investment decisions. First, the tested generator understates volatility and tail losses and distorts cross-asset dependence. Second, these discrepancies coexist with changes in allocation behavior: replacement exhibits frequent near-zero lower-tail means and substantial sensitivity to the modified denominator, while some generation errors are associated with concentration or turnover. Third, large in-window ratios do not reliably translate into higher net returns or lower realized tail losses. These are jointly observed model-risk features, not a demonstrated causal chain from a particular generation error to underperformance.

The resampling comparison also prevents a one-dimensional interpretation of fidelity. TimeGAN has higher mean net returns than IID and moving-block controls at each tested mass, despite its documented statistical errors. All direct return, Sharpe and realized-RR intervals nevertheless include zero. Thus, poor fidelity does not establish uniformly inferior decisions, and the observed mean ordering does not establish superiority over simpler generators. Evaluating scenarios only by draw count, reconstruction loss or attained optimization score can miss their interaction with the allocation criterion, concentration, turnover and transaction costs.

\begin{figure}[!htbp]
\centering
\includegraphics[width=.96\textwidth,height=.58\textheight,keepaspectratio]{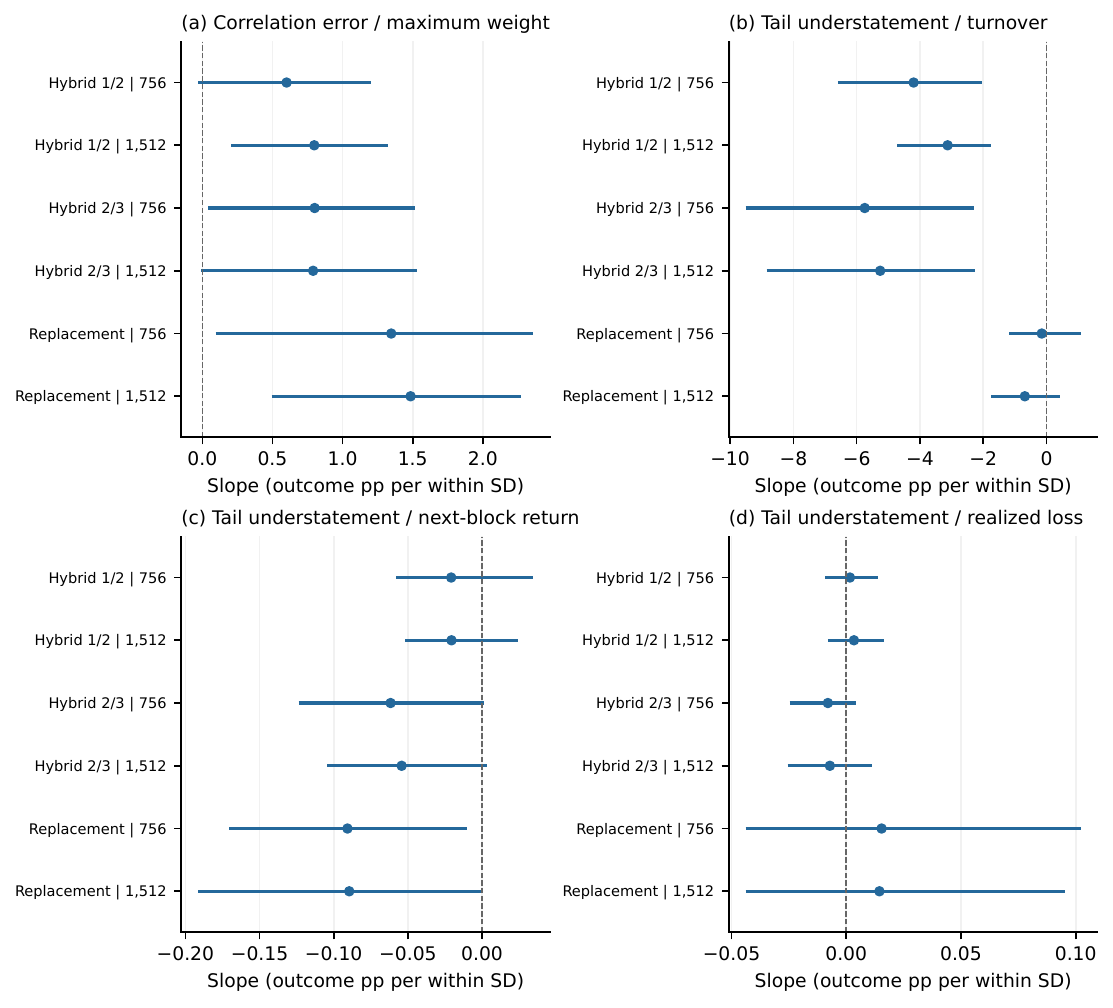}
\caption{Exploratory generation-error associations, with four separate two-way fixed-effect regressions per configuration. Points are outcome percentage-point changes per one original within-panel predictor standard deviation. Bars are unadjusted 95\% circular-block bootstrap intervals: 1,999 draws, 38 decision dates per block, synchronized across five seeds; each configuration has 990 decisions on 198 shared dates. Return contrast is next-block compounded net return minus historical RRP. Realized loss uses fractional 5\% tails within only 15 or 20 daily observations. These associations do not identify causal transmission; alternative six- and 12-decision blocks remain in the electronic results.}\label{fig:generation_policy_associations}
\end{figure}

To examine the proposed distribution--decision link beyond selected diagnostic windows, Figure~\ref{fig:generation_policy_associations} joins the main experiment's saved allocation draw to all 198 decision dates, five seeds and six mass--count configurations. Each configuration supplies a balanced panel of 990 decisions on one shared market history. Four separate regressions remove decision-date and seed means. Coefficients express the outcome change, in percentage points, per one standard deviation of the remaining predictor variation. Correlation error is the mean absolute pairwise error; tail understatement is the negative signed asset-level loss-ETL error. These are explanatory summaries of previously observed policies, not predictive models selected on an untouched evaluation sample.

Figure~\ref{fig:generation_policy_associations} links greater correlation error to higher maximum target weights, with slopes of 0.60--1.49 percentage points across configurations. Greater tail understatement instead accompanies lower hybrid turnover, with slopes from -3.13 to -5.75 percentage points; the replacement-turnover intervals include zero. This direction rules out treating a monotone error--turnover channel as an established explanation of the strategy-level differences. Subsequent-return slopes are small, and interval exclusion is sensitive to block length: the replacement configuration with 756 draws excludes zero with 38-decision blocks but not with six- or 12-decision blocks. All realized-loss intervals include zero. Neither these unadjusted, exploratory associations nor the denominator sensitivity identify a causal mediation of investment performance. Short holding-block loss tails and heavily overlapping training windows further limit precision; the longest block spans approximately one training window but cannot ensure interval coverage.

The count--mass comparison, conditional intervals and fixed-path cost repricing in Sections~\ref{sec:investment}, \ref{sec:inference} and \ref{sec:costs_episodes} address distinct questions. More draws refine sampling from a fitted distribution; fixed-target repricing measures fee exposure rather than transaction-cost-aware allocation. They do not establish invariance to the 756-day training window or 20-day decision interval. The full-period checks in Sections~\ref{sec:floor_sensitivity} and~\ref{sec:resampling} remove the ten-window selection restriction, but source comparisons still fix the added count at 756 and reuse the same fitted models and historical path. Limited longer-training diagnostics in Section~\ref{sec:training_variation} do not settle training-budget effects throughout the experiment.

The findings concern one retrospectively selected ETF universe, one historical interval, a particular GRU TimeGAN implementation and fixed training effort. They do not establish a result for all TimeGAN implementations or other generators. Possible local numerical solutions, the risk-free proxy conversion and incomplete independent verification of adjusted market prices remain relevant. A favorable in-window ratio is not interchangeable with realized tail protection or net wealth growth.

The evaluation evolved during reexamination and was not preregistered; there is no untouched outer evaluation. Seeds represent alternative fitted policies on a shared history, not independent market samples. Both the original performance intervals and the new panel intervals condition on saved fits and policies: they omit retraining, reoptimization and design selection. The panel intervals are pointwise rather than multiplicity-adjusted. Intervals containing zero establish neither equivalence nor absence of a relationship. These boundaries motivate joint assessment of scenario fidelity, numerical reliability and realized outcomes.

\FloatBarrier

\section{Conclusion}\label{sec:discussion_conclusion}

The main 31-policy experiment does not establish a consistent net-performance advantage from replacing or supplementing historical returns with the fitted TimeGAN. Higher-mass hybrid configurations have mean annualized net returns of 3.99--4.25\%, compared with 3.48\% for historical allocation, while equal-mass hybrids earn about 2.80\%. These descriptive differences vary across seeds, and all conditional generated-versus-historical return intervals include zero at the reference block length. Passive references deliver greater net growth in this sample, while the historical strategy retains a higher realized tail ratio than buy-and-hold.

The full-period controls extend the evidence beyond selected diagnostic windows. TimeGAN has higher mean returns than IID and moving-block resampling at each tested mass, but all six direct return, Sharpe and realized-RR intervals include zero under each reported block length. Full replacement also has greater daily tail loss than the resampling references. Changing to a historical-scale denominator relocates 35.21\% of replacement target weight on average and sharply lowers the attained objective, while the mean net-return change has a zero-containing interval and daily loss-CVaR rises. These findings distinguish criterion sensitivity from evidence of better investment outcomes; they neither establish equivalence between sources nor justify choosing a floor from its realized return.

This study evaluates a specified generator rather than the best attainable TimeGAN; the fixed budget and limited longer-training diagnostics do not establish optimal model adequacy. The comparison evaluates the reliability of learned scenarios and their interaction with a tail-sensitive decision rule. It separates replacement from supplementation and synthetic probability mass from generated sample size under common allocation and execution rules. The results underscore the distinction between increasing the number of generated scenarios and improving their usefulness for portfolio decisions. Full-panel exploratory associations provide evidence of links to concentration and hybrid turnover, but do not identify a stable subsequent-return channel or a causal decomposition. The findings are specific to the tested universe, period, generator, and empirical objective; the conditional intervals do not incorporate model refitting or policy reoptimization. They support joint evaluation of scenario fidelity, numerical reliability, and investment outcomes when assessing synthetic data for portfolio decisions.

\section*{Acknowledgement}
The author thanks Prof.~Svetlozar Rachev of the Department of Mathematics \& Statistics, Texas Tech University, for his helpful comments and suggestions on this work.

\FloatBarrier
\clearpage
\appendix
\setcounter{table}{0}
\setcounter{figure}{0}
\renewcommand{\thetable}{A\arabic{table}}
\renewcommand{\thefigure}{A\arabic{figure}}
\renewcommand{\theHtable}{appendixA.\arabic{table}}
\renewcommand{\theHfigure}{appendixA.\arabic{figure}}
\section{Auxiliary Diagnostics}\label{sec:appendix_results}
This appendix contains auxiliary diagnostics and interval sensitivity results not displayed in the main text. Tables already replaced by main-text statistical figures are not repeated. The two training tables retain only information omitted from the paired reconstruction/correlation plot. All records come from the completed experiments; no model fitting, optimization or bootstrap experiment is repeated for this presentation.

\Needspace{9\baselineskip}
\subsection{Asset Distribution Diagnostics}
\begin{figure}[!htbp]
\centering
\includegraphics[width=.98\textwidth,height=.70\textheight,keepaspectratio]{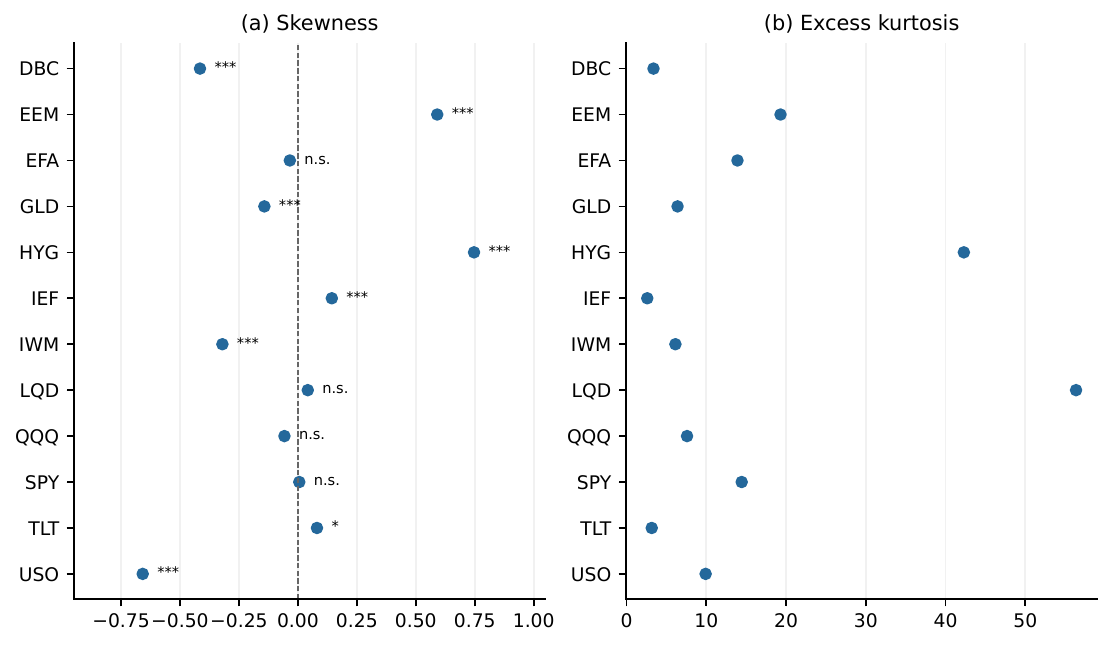}
\caption{Skewness and excess kurtosis of 4,712 daily returns for each ETF. Skewness markers retain the original annotations: *** $p<0.001$, * $p<0.05$ and n.s. $p\geq0.05$. All kurtosis and joint normality $p$-values are below 0.001. Conventional significance assumes independent observations; dependence and conditional heteroskedasticity limit that interpretation. The plot is descriptive and does not validate a specific allocation criterion.}\label{fig:supp_distribution}
\end{figure}

Figure~\ref{fig:supp_distribution} provides the auxiliary comparison. All assets have positive excess kurtosis, while skewness differs in sign and magnitude. The annotations should be read with the dependence caveat.

\FloatBarrier

\Needspace{9\baselineskip}
\subsection{Extreme-Date Market and NAV Comparisons}
\begin{figure}[!htbp]
\centering
\includegraphics[width=.98\textwidth,height=.70\textheight,keepaspectratio]{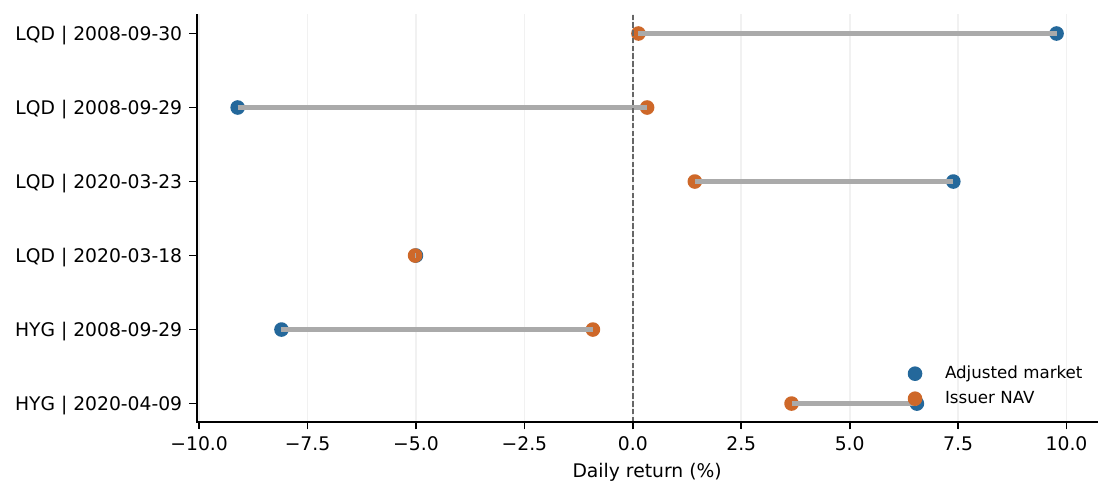}
\caption{Paired adjusted-market and distribution-inclusive issuer NAV daily returns at the six reported ETF--date cases. Dates identify the six reported issuer cross-checks. The comparisons diagnose differences between market prices and NAV, not verified data errors or replacement observations.}\label{fig:supp_market_nav}
\end{figure}

Figure~\ref{fig:supp_market_nav} provides the auxiliary comparison. Market and NAV returns differ sharply for several LQD and HYG dates; the LQD observation on March 18, 2020 is much closer across valuation bases.

\FloatBarrier

\Needspace{9\baselineskip}
\subsection{Limited Training Diagnostics}
\begin{figure}[!htbp]
\centering
\includegraphics[width=.98\textwidth,height=.70\textheight,keepaspectratio]{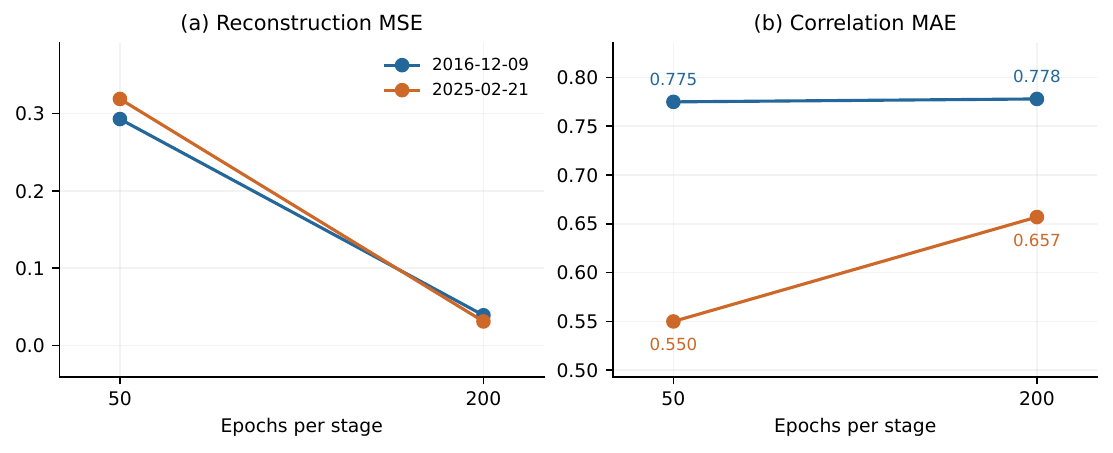}
\caption{Existing primary-seed 50/200-epoch comparisons in two representative windows. Epochs are per training stage. Reconstruction MSE falls, whereas correlation MAE does not improve. These are limited paired diagnostics, not evidence that 50 epochs are sufficient or that further tuning cannot improve generation.}\label{fig:supp_training_pairs}
\end{figure}

Figure~\ref{fig:supp_training_pairs} provides the auxiliary comparison. Both paired windows show lower reconstruction MSE at 200 epochs but slightly or substantially greater correlation MAE. The other reported windows have only the recorded 50-epoch diagnostics; the table retains them.

Table~\ref{tab:appendix_other_training} retains the eight other representative-window records for which only the recorded 50-epoch diagnostics are available. Table~\ref{tab:appendix_paired_marginals} gives the marginal volatility and loss errors omitted from the paired MSE/correlation plot. These auxiliary records do not establish full-period convergence or sufficient training.

\begin{table}[!htbp]
\centering\small\setlength{\tabcolsep}{3pt}
\caption{Additional representative-window training diagnostics not shown in the paired plot}\label{tab:appendix_other_training}
\begin{tabularx}{\textwidth}{@{}lXrrrrr@{}}
\toprule
Seed & Information end & Epochs & MSE & Corr. MAE & $\Delta\sigma$ & $\Delta L$ \\
\midrule
23456 & 2015-01-15 & 50 & 0.318 & 0.580 & -0.310 & -0.988 \\
23456 & 2017-11-22 & 50 & 0.313 & 0.766 & -0.384 & -1.169 \\
34567 & 2010-04-12 & 50 & 0.440 & 0.574 & -0.559 & -1.824 \\
34567 & 2018-12-06 & 50 & 0.313 & 0.750 & -0.187 & -0.938 \\
45678 & 2011-01-25 & 50 & 0.390 & 0.572 & -0.516 & -1.856 \\
45678 & 2018-09-11 & 50 & 0.333 & 0.738 & -0.418 & -1.140 \\
56789 & 2015-02-13 & 50 & 0.387 & 0.606 & -0.556 & -1.358 \\
56789 & 2018-05-17 & 50 & 0.302 & 0.761 & -0.286 & -1.011 \\
\midrule
\multicolumn{7}{@{}p{\textwidth}@{}}{\footnotesize\textit{Note:} Epochs are per stage. These rows have no matched 200-epoch observation. MSE and correlation MAE are unitless; $\Delta\sigma$ and $\Delta L$ are generated-minus-historical daily percentage-point errors in marginal volatility and integer-ceiling 5\% tail loss. They are local diagnostics, not independent market replications.} \\
\bottomrule
\end{tabularx}
\end{table}

\begin{table}[!htbp]
\centering\small
\caption{Marginal errors in the two existing primary-seed training pairs}\label{tab:appendix_paired_marginals}
\begin{tabular}{lrrr}
\toprule
Information end & Epochs per stage & $\Delta\sigma$ (pp/day) & $\Delta L$ (pp/day) \\
\midrule
2016-12-09 & 50 & -0.213 & -0.972 \\
2016-12-09 & 200 & 0.019 & -0.486 \\
2025-02-21 & 50 & -0.375 & -1.243 \\
2025-02-21 & 200 & 0.004 & -0.557 \\
\midrule
\multicolumn{4}{l}{\footnotesize\textit{Note:} Seed 12345. Errors are generated minus historical.} \\
\bottomrule
\end{tabular}
\end{table}

\FloatBarrier

\Needspace{9\baselineskip}
\subsection{First-Window Generation Variability}
\begin{figure}[!htbp]
\centering
\includegraphics[width=.98\textwidth,height=.70\textheight,keepaspectratio]{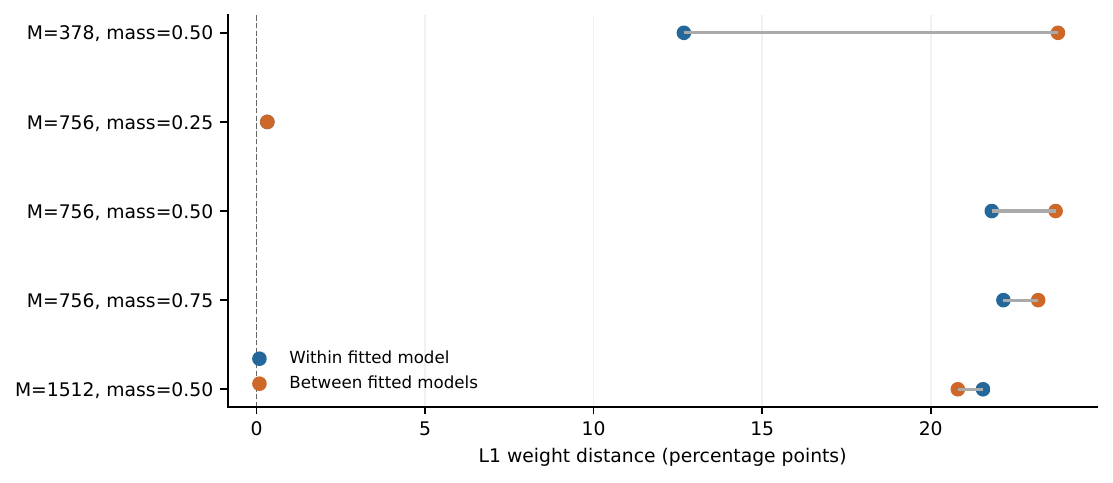}
\caption{First-window allocation variation from independent generation. Within-model L1 distances average three draws relative to their own fitted-model mean; between-model distances compare five model means with their common mean. Values are percentage points. All use the recorded fractional-tail search with two random starts and at most 300 iterations. These are focused diagnostics, not full-period performance or globally certified solutions.}\label{fig:supp_generation_variation}
\end{figure}

Figure~\ref{fig:supp_generation_variation} provides the auxiliary comparison. At synthetic mass one-half, increasing draw count from 756 to 1,512 leaves within-model allocation variation near 22 percentage points; the between-model distance declines modestly.

\FloatBarrier

\Needspace{9\baselineskip}
\subsection{Illustrative Asset Allocations}
\begin{figure}[!htbp]
\centering
\includegraphics[width=.98\textwidth,height=.70\textheight,keepaspectratio]{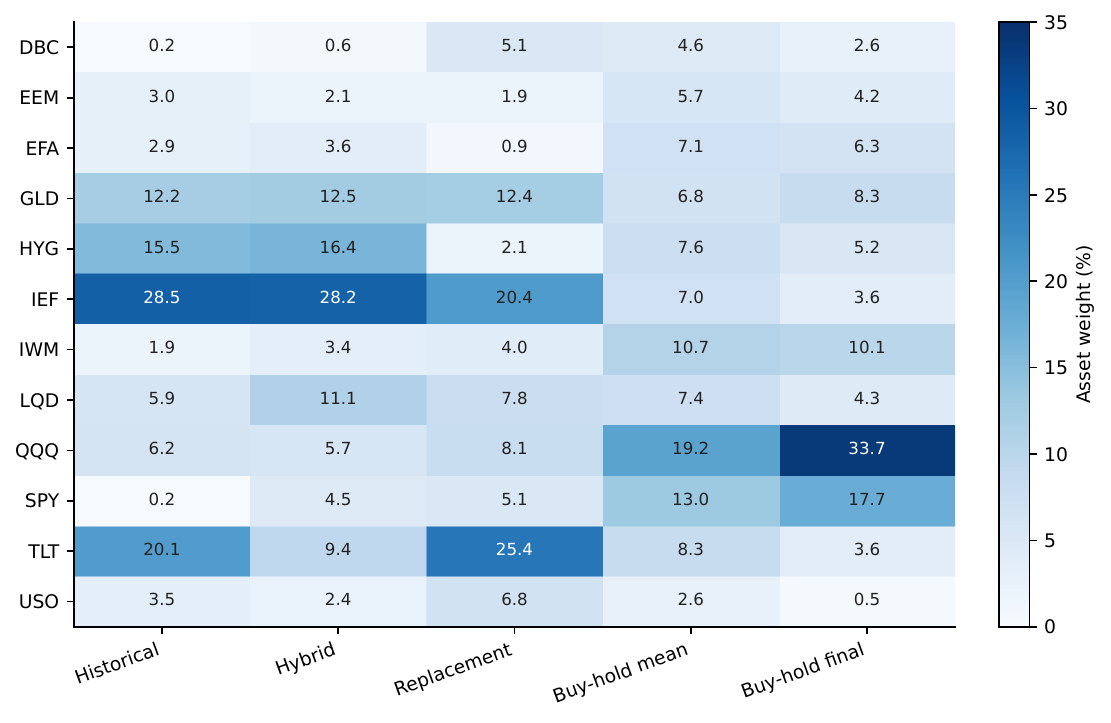}
\caption{Average daily held ETF weights and buy-and-hold final weights, in percent. Generated policies use seed 12345, 1,512 rows and hybrid mass $2/3$, selected independently of investment performance. The first four columns are time means; the final column is a terminal observation and is not a mean.}\label{fig:supp_allocation}
\end{figure}

Figure~\ref{fig:supp_allocation} provides the auxiliary comparison. Historical and hybrid mean exposures favor Treasury and credit ETFs; buy-and-hold drifts toward QQQ and SPY by the final date. Mean and ending weights describe different quantities.

\FloatBarrier

\Needspace{9\baselineskip}
\subsection{Passive-Reference Interval Sensitivity}
\begin{figure}[!htbp]
\centering
\includegraphics[width=.98\textwidth,height=.70\textheight,keepaspectratio]{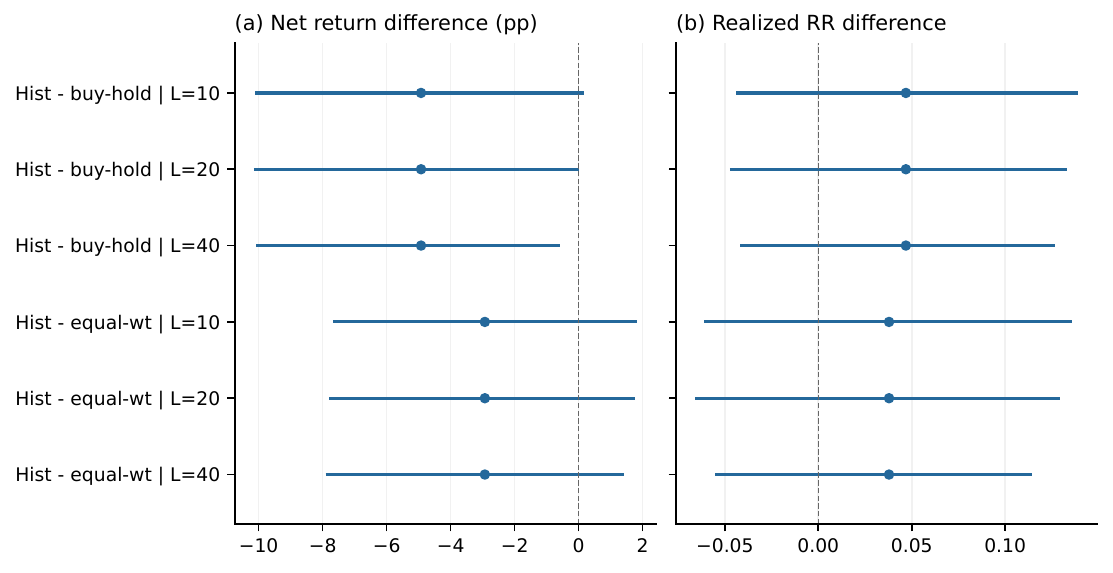}
\caption{Historical-minus-passive net-return and realized-RR contrasts with pointwise conditional 95\% intervals at daily block lengths 10, 20 and 40. Buy-and-hold and equal-weight are distinct references. All 2,000 calendar-block resamples are synchronized across policies. Return differences are annualized percentage points; RR is unitless. The buy-and-hold return upper endpoint at length 20 is approximately $-3.76\times10^{-4}$ annualized percentage points; its proximity to zero limits the precision of a graphical reading.}\label{fig:supp_passive}
\end{figure}

Figure~\ref{fig:supp_passive} provides the auxiliary comparison. Historical RRP has lower point-estimate net growth than both passive references. Interval exclusion for the buy-and-hold return contrast depends on block length, while both passive-reference RR intervals include zero at every reported length.

\FloatBarrier

\Needspace{9\baselineskip}
\subsection{Fixed-Calendar Episodes}
\begin{figure}[!htbp]
\centering
\includegraphics[width=.98\textwidth,height=.70\textheight,keepaspectratio]{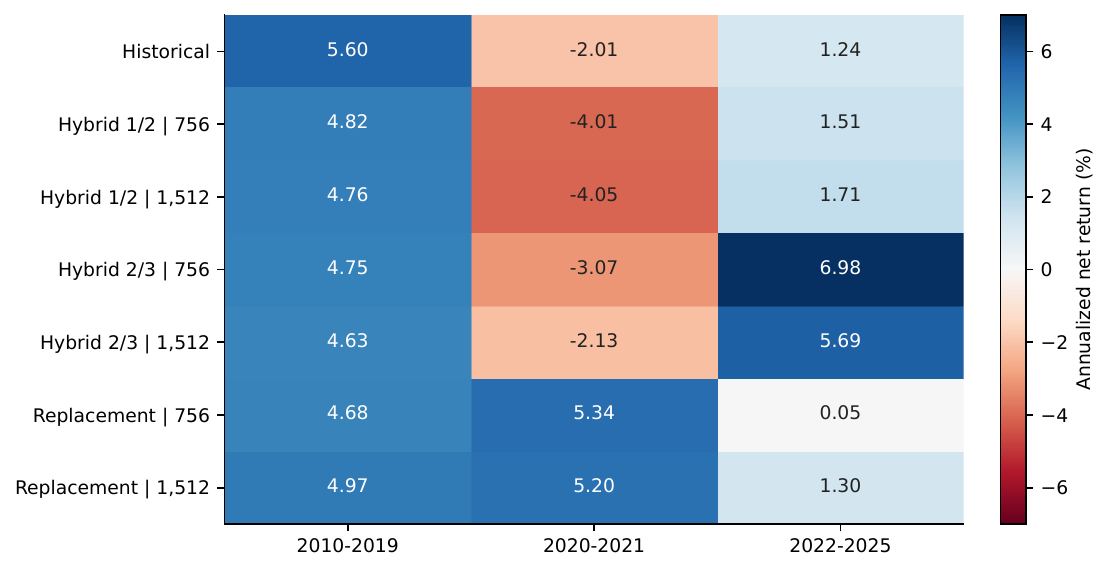}
\caption{Annualized net returns within three fixed calendar slices of the same saved policies. Generated cells average five fixed-seed statistics. Colors are centered at zero; values are percentages. These are descriptive episodes rather than independent market replications or newly fitted period-specific policies.}\label{fig:supp_episodes}
\end{figure}

Figure~\ref{fig:supp_episodes} provides the auxiliary comparison. Policy rankings change across calendar slices: replacement is positive in 2020--2021 while the historical and hybrid means are negative, and higher-mass hybrids are stronger in 2022--2025.

\FloatBarrier

\Needspace{9\baselineskip}
\subsection{Additional Block-Length Sensitivity}
\begin{figure}[!htbp]
\centering
\includegraphics[width=.98\textwidth,height=.70\textheight,keepaspectratio]{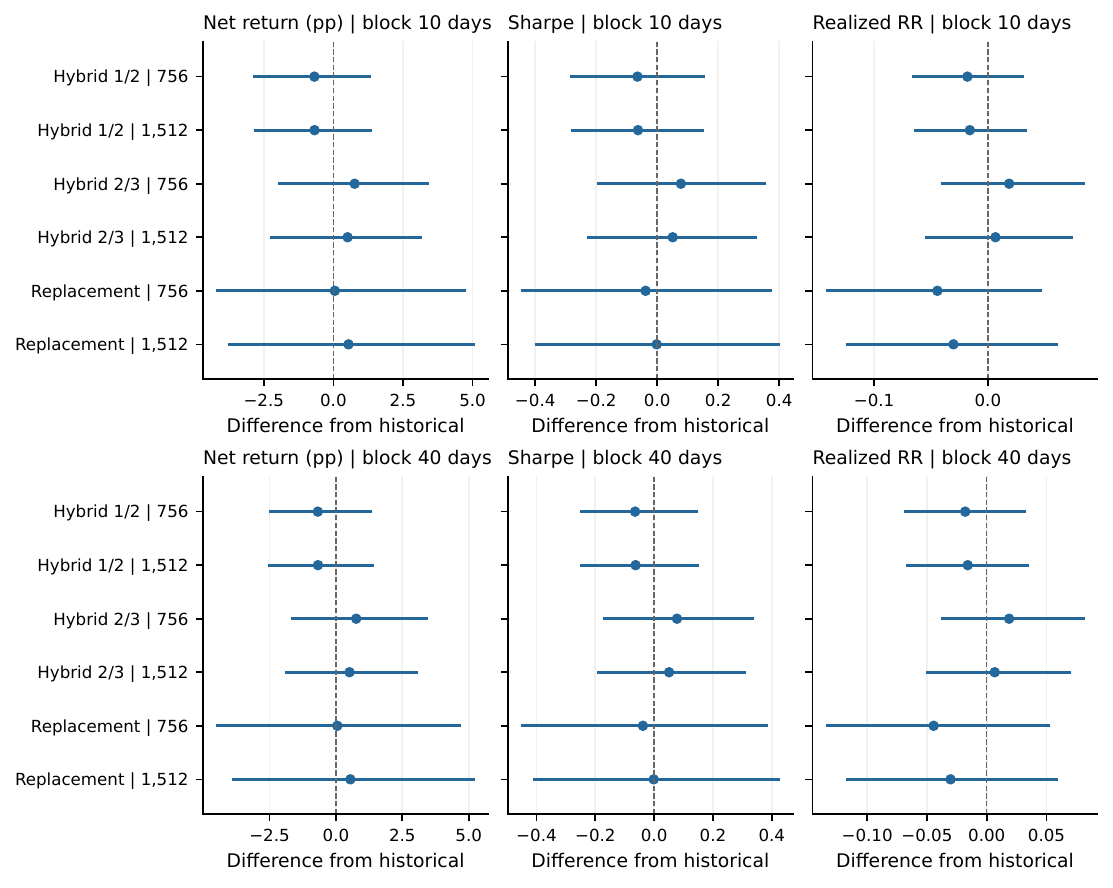}
\caption{Conditional configuration-average differences from historical RRP at daily block lengths 10 and 40. Columns show annualized return differences in percentage points, Sharpe differences and realized fractional-RR differences. Bars are pointwise 95\% percentile intervals from the existing 2,000 synchronized overlapping-block resamples; each cell averages the five fixed-seed statistics. These auxiliary panels complement the main 20-day figure and omit refitting, reoptimization and specification-selection uncertainty.}\label{fig:appendix_block_sensitivity}
\end{figure}

Figure~\ref{fig:appendix_block_sensitivity} supplies the 10- and 40-day intervals omitted from the main 20-day reference plot. All configuration-average return, Sharpe and realized-RR intervals include zero at both alternative lengths. Interval endpoints vary, especially for replacement. These are conditional sensitivity checks on the same fixed paths, not independent confirmations or equivalence tests.

\FloatBarrier

\bibliographystyle{apalike}
\bibliography{paper}
\end{document}